\documentclass[pdftex, colorlinks=true, linkcolor=blue, citecolor=blue, urlcolor=blue]{aa}

\usepackage[mathscr]{euscript}
\usepackage{comment}
\usepackage{color}
\usepackage{hyperref}
\usepackage{lastpage}
\usepackage[normalem]{ulem} 
\usepackage{float}
\usepackage[graphicx]{realboxes}

\usepackage{txfonts,xcolor}
\usepackage{colortbl}
\usepackage{placeins}
\usepackage{multicol}

\definecolor{indigo}{rgb}{0.6, 0.4, 0.42}
\definecolor{forestgreen}{rgb}{0.13, 0.55, 0.13}

\newcommand{\St}{\text{St}}
\newcommand{\Sc}{\text{Sc}}

\begin{document}

\title{Turbulence in protoplanetary disks: A systematic analysis of dust settling in 33 disks}

%
\author{Marion~Villenave\inst{1}
        \and Giovanni P. Rosotti\inst{1}
        \and Michiel Lambrechts\inst{2}
        \and Alexandros Ziampras\inst{3,4}
        \and Christophe Pinte\inst{5}
        \and Fran\c cois M\'enard\inst{5}
        \and Karl R. Stapelfeldt\inst{6}
        \and Gaspard Duch\^ene\inst{5,7}
        \and Emily Baylock\inst{8}
        \and Kiyoaki Doi\inst{9}
    }
    
\institute{Universit\'a degli Studi di Milano, Dipartimento di Fisica, via Celoria 16, 20133 Milano, Italy
\email{marion.villenave@univ-grenoble-alpes.fr}
\and Centre for Star and Planet Formation, Globe Institute, University of Copenhagen, \O ster Voldgade 5-7, 1350 Copenhagen, Denmark
\and 
Astronomy Unit, School of Physics and Astronomy, Queen Mary University of London, London E1 4NS, UK
\and
Ludwig-Maximilians-Universit{\"a}t M{\"u}nchen, Universit{\"a}ts-Sternwarte, Scheinerstr.~1, 81679 M{\"u}nchen, Germany
\and 
Univ. Grenoble Alpes, CNRS, IPAG, 38000 Grenoble, France 
\and
Jet Propulsion Laboratory, California Institute of Technology, 4800 Oak Grove Drive, Pasadena, CA 91109, USA
\and 
Astronomy Department, University of California Berkeley, Berkeley CA 94720-3411, USA
\and
California Institute of Technology, Pasadena, CA 91125, USA
\and
Max-Planck Institute for Astronomy, Königstuhl 17, D-69117 Heidelberg, Germany
}

\abstract
{The level of dust vertical settling and radial dust concentration in protoplanetary disks is of critical importance for understanding the efficiency of planet formation. Here, we present the first uniform analysis of the vertical extent of millimeter dust for a representative sample of 33  protoplanetary disks, covering broad ranges of disk evolutionary stages and stellar masses. 
We used radiative transfer modeling of archival high-angular-resolution ($\lesssim0.1\arcsec$) ALMA dust observations of inclined and ringed disks to estimate their vertical dust scale height, which was compared to estimated gas scale heights to characterize the level of vertical sedimentation. 
In all 23 systems for which constraints could be obtained, we find that the outer parts of the disks are vertically settled. 
Five disks allow for the characterization of the dust scale height both within and outside approximately half the dust disk radius, showing a lower limit on their dust heights at smaller radii. This implies that the ratio between vertical turbulence, $\alpha_z$, and the Stokes number, $\alpha_z/\St$, decreases radially in these sources. For 21 rings in 15 disks, we also constrained the level of radial concentration of the dust, finding that about half of the rings are compatible with strong radial trapping. In most of these rings, vertical turbulence is found to be comparable to or weaker than radial turbulence,
which is incompatible with the turbulence generated by the vertical shear instability at these locations. We further used our dust settling constraints to estimate the turbulence level under the assumption that the dust size is limited by fragmentation, finding typical upper limits around $\alpha_\text{frag}\lesssim10^{-3}$.  In a few sources, we find that turbulence cannot be the main source of accretion.
Finally, in the context of pebble accretion, we identify several disk regions that have upper limits on their dust concentration 
that would allow core formation to proceed efficiently, even at wide orbital distances outside of 50\ au. 
}
\maketitle

\section{Introduction} \label{sec:intro}

Protoplanetary disks are the birthplaces of planets. As such, understanding disk properties is critical to determining how and where planets are forming. Two main theories can explain planet formation. On the one hand, dense and cold disks can form massive planets directly from the available dust and gas via the gravitational instability process. This is similar to the star formation mechanism and is expected to create planets at large separations from their host star ($>100$au). On the other hand, the core accretion scenario requires the progressive growth of initially submicron-sized particles to form larger bodies and eventually the rocky cores of giant planets. In this scenario, an important concentration of solids is necessary to accelerate the dust growth. Indeed, the streaming instability~\citep{Youdin_2005,Johansen_Youdin_2007}, currently one of the favored mechanisms allowing the sufficiently rapid growth from pebbles to planetesimals, can develop if the dust-to-gas ratio is high, of order unity. After the planetesimal formation stage,  growth of planetary cores via pebble accretion~\citep{Lambrechts_2012} is also enhanced if the layer of pebbles is vertically thin. 

The spatial distribution of dust and its enhancement with respect to the gas is determined by dust--gas interactions. As dust orbits the central star, gas drag causes it to drift radially toward a local pressure maximum. In the vertical direction, initial oscillations of dust particles around the disk midplane are dampened due to gas drag, which leads dust to fall vertically toward the disk midplane.  
The level of dust concentration is directly related to the ratio $\alpha/\St$, where $\St$ is the Stokes number that parametrizes the coupling of dust with gas, and $\alpha$ quantifies the turbulence level in the disk. High values of this ratio imply strong mixing of dust with gas, while low values ($\alpha/\St<<1$) mean that the dust is decoupled from the gas and concentrated into geometrically thin regions. 
Despite their critical role in planet formation, observational constraints on the level of gas turbulence and the coupling between dust and gas remain limited. 

In order to explain the accretion of the disk toward the star, protoplanetary disks are predicted to be turbulent. Various hydrodynamical or magnetohydrodynamic processes can be at the origin of the physical gas turbulence in disks~\citep[see][for a review]{Lesur_2023} and still need to be constrained by observations. Both the strength and isotropy of the predicted turbulence can be compared with observations. For example, mechanisms such as the vertical shear instability \citep[VSI,][]{Nelson_etal_2013} imply very strong vertical turbulence associated with weak radial mixing~\citep[e.g.,][]{Stoll_2017}. On the contrary, gravitational instability results in the opposite anisotropy of turbulence~\citep[e.g.,][]{Bethune_etal_2021}, while mechanisms such as the streaming instability, embedded planets, or ambipolar diffusion imply turbulence of similar strength in both the vertical and radial directions~\citep[e.g.,][]{Youdin_2005, Simon_etal_2013}. 
Finally, alternate views can also explain accretion without requiring turbulence, as in the case of MHD winds for example~\citep[e.g.,][]{Bai_Stone_2011, Tabone_2022, Lesur_2023}.

Measuring the strength of turbulence is notoriously difficult~\citep[see][for a recent review]{Rosotti_2023}, in part because turbulence is predicted to be very weak~($\alpha\ll1$). Direct measurements via line broadening have been performed in about half a dozen systems~\citep{Flaherty_2015, Flaherty_2018, Flaherty_2020, Teague_2016}, typically finding that weak turbulence, $\alpha \lesssim 10^{-3}$, may be common (although there are exceptions, see e.g., \citealt{PanequeCarreno_2024, Flaherty_2024}). These studies are, however, limited to a few objects, due to the high quality of the observations required, and only allow for the measurement of a global value for the turbulence strength at the disk surface. On the other hand, the radial and vertical concentration of dust can allow us to go beyond these limitations and evaluate the variation and anisotropy of turbulence throughout the disk and for a larger number of objects. Dust accumulation also traces the  turbulence at the disk midplane, which is important for dust growth.

Thanks to ALMA, several dozen of protoplanetary disks have been observed at high angular resolution ($\lesssim0.1''$). These observations have revealed a broad range of substructures in millimeter continuum, with annular rings being the most common~\citep{Andrews_2018, Long_2018}. Such observations can be used to directly observe and characterize the radial and vertical structure of protoplanetary disks, and obtain insights into the turbulence levels and structure in the disks.  High-angular-resolution observations of edge-on disks are the most direct way to retrieve information on their vertical structure~\citep{Villenave_2020, Villenave_2022, Lin_2021, Lin_2023, sturm_2023}. However, such systems are relatively rare~\citep[$\sim 20$ -- 30 are currently known,][]{Angelo_2023}, and do not provide a good view of their radial structure. On the other hand,  systems at intermediate inclinations that possess clear ring and gap features are favorable targets to constrain both the vertical~\citep{Pinte_2016, Doi_Kataoka_2021, Pizzati_2023} and radial~\citep{Dullemond_2018, Rosotti_2023} concentration of dust particles.  
Here, we propose to expand the analysis of the vertical and radial concentration of dust to a larger sample and to systematically look for potential radial variations in the turbulence strength within the disks. Our goal is also to characterize the turbulence level and its potential anisotropy in a large sample of protoplanetary disks, and to discuss the implications regarding the origin of the turbulence and wide-orbit planet formation.

We present the analysis of archival ALMA continuum observations of 33 inclined and ringed protoplanetary disks. 
The sample selection and data reduction are discussed in Sect.~\ref{sec:obs}. Then, Sect.~\ref{sec:model} describes the methodology employed to model the disks. Sect.~\ref{sec:results} introduces the results for the vertical concentration of dust in the disks. In Sect.~\ref{sec:constraintsturbulence} we determine the level of vertical and radial dust--gas coupling ($\alpha/\St$) in the different rings, and discuss the anisotropy of turbulence and implication regarding its possible origin. In Sect.~\ref{sec:discussion}, we break the degeneracy between $\alpha$ and $\St$ by assuming that the dust size is limited by fragmentation. We discuss the results in the context of disk accretion and potential for planet formation.  We conclude with a discussion on the implications of our results for wide-orbit planet formation via pebble accretion (Sect.~\ref{sec:impplan}) and our final remarks (Sect.~\ref{sec:conclusion}).

\section{Observations}
\label{sec:obs}
\subsection{Sample}

We selected disks with moderate to high inclination that present ring structures and for which good quality ALMA data at band~6~\citep[$\sim1.3\,\text{mm}$;][]{Ediss_2004, Kerr_2014} or 7~\citep[$\sim0.9$\,mm;][]{Mahieu_2012} are available in the archive. We report observation characteristics of the sources in Appendix~\ref{app:observations}.

We analyzed a total of 33 disks. Our sample includes 15 disks from the DSHARP survey~\citep[AS209, DoAr25, DoAr33, Elias20, Elias24, Elias27, GWLup, HD142666, HD143006, HD163296, MYLup, SR4, Sz114, WaOph6, WSB52,][]{Andrews_2018}, for most of which a measurement of the height with a similar technique was previously obtained or attempted by \citet{Pizzati_2023}. We also include five disks from the Taurus survey by \citet[][CI Tau, DL Tau, DS Tau, GO Tau, MWC 480]{Long_2018} and one from the ODISEA survey \citep[ISO-Oph 17, ][]{Cieza_2021}. Finally, the rest of the sample constitutes of 12 disks previously published in individual papers~\citep[AA Tau, DM Tau, GM Aur, Haro6-5B, HL Tau, J1608, J1610, LkCa 15, Oph163131, PDS70, RYTau, V1094Sco;][]{Loomis_2017, Hashimoto_2021, Huang_2020, Stephens_2023, Facchini_2020, Villenave_2020,  Villenave_2022, Benisty_2021, Ribas_2024, vanTerwisga_2018}.

\autoref{tab:disk_parameters} gathers the disk and stellar parameters considered for this study. Specifically, we compile the stellar masses and radii, which are input parameters required for the modeling presented in Sect.~\ref{sec:model}.  For most disks, we estimate the stellar radius based on the published stellar luminosity following    $   R_\star = \sqrt{L_\star / 4 \pi \sigma T_\text{eff}^4}$ (however, see Sect.~\ref{sec:fittingSurfaceDensity} for systems marked with $^*$ in $R_\star$ column of \autoref{tab:disk_parameters}). The sample covers a wide range of stellar masses, between 0.36 M$_\sun$ and 2.04 M$_\sun$. It also includes objects with different accretion rates spanning four orders of magnitudes, which could be at different evolutionary stages within the Class~II phase~\citep[although young stellar object accretion is episodic,][]{Evans_2009}. 

\subsection{Data reduction}
 
\begin{figure*}
    \centering
    \includegraphics[width = 1\textwidth]{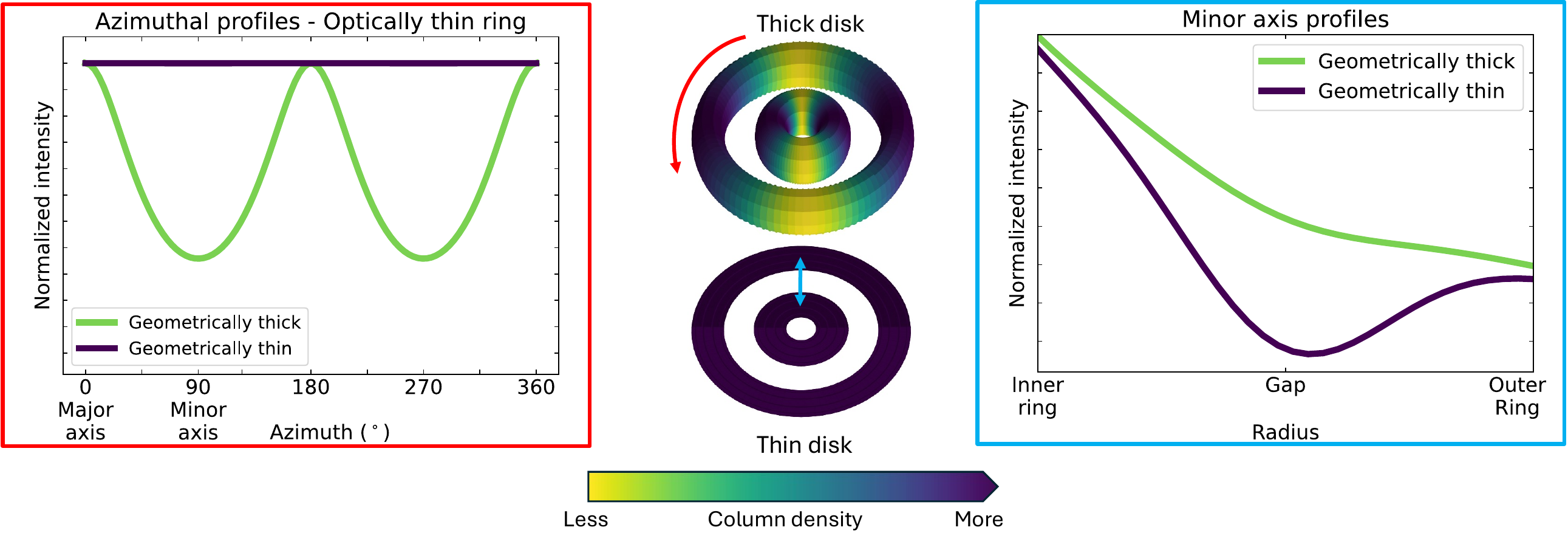}
    \caption{Schematic representation of the expected effects of disk thickness. \emph{Left:} Effect of ring vertical thickness on the azimuthal profiles of a radially narrow, optically thin ring.  A vertically thin ring (dark) shows no variation in its azimuthal profile along the ring, while a thick ring (green) displays strong azimuthal variation along the ring. \emph{Right:} Effect of disk vertical thickness on the minor axis brightness of a disk at the location of a gap (right). A vertically thin gapped disk (dark) shows deep gap along the minor axis, while a thick gapped disk (green) displays shallow gap along the minor axis direction.}
    \label{fig:schematic}
\end{figure*}

For each object, we gathered calibrated visibility files either from the ALMA archive or directly from the first publisher team. Data collected from first publisher teams are the DSHARP targets~\citep{Andrews_2018}, DM Tau~\citep{Hashimoto_2021}, Haro6-5B~\citep{Villenave_2020}, Oph163131~\citep{Villenave_2022}, PDS 70~\citep{Benisty_2021}, RY Tau~\citep{Ribas_2024}, and some Taurus disks from \citet[][DL Tau, DS Tau, GO Tau, MWC 758, see Table~\ref{tab:observations} and Table~\ref{tab:observations_no_constraints}]{Long_2018}. Those datasets had been previously self-calibrated.
For datasets collected from the archive, we employed the automated self-calibration script developed at  NRAO\footnote{\url{https://science.nrao.edu/srdp/self-calibration-preview}, \url{https://github.com/jjtobin/auto_selfcal/}}, which was successful to improve the data quality of V1094 Sco and ISO-Oph 17. 
Then, for each disk, we generated an image using the \texttt{tclean} algorithm, using CASA version 5.1.1-5, consistently with the version used to generate the DSHARP images. All images were generated with the spectral model \texttt{mfs} and \texttt{nterms=1}. A multi-scale synthesis was used for all disks and the images were cleaned up to $\sim5\sigma$. We used a briggs weighting, with a \texttt{robust} parameter between -0.5 and 0.5 as is indicated in Table~\ref{tab:observations} and Table~\ref{tab:observations_no_constraints}. We also report the root mean square (rms) of the final images in those tables. All the reduction parameters, the visibilities, and cleaned disk images have been uploaded on zenodo (\href{https://doi.org/10.5281/zenodo.14054952}{zenodo.14054952}).
Finally, we generate a deprojected image based on the inclinations reported in \autoref{tab:disk_parameters}.

\section{Radiative transfer modeling}
\label{sec:model}

In this section, we first describe the expected effect of vertical thickness to the appearance of a protoplanetary disk. Then, we detail the modeling procedure of this work, which allows us to fit each disk and obtain constraints on their vertical extent. We follow the approach that \cite{Pinte_2016} developed for the case of HL Tau, and which was also used by  \cite{Pizzati_2023} on 15 DSHARP disks. Compared to the study of \cite{Pizzati_2023}, we now consistently compute  the temperature structure of the disk, consider a full grain size distribution, and analyze the disk emission  not only along the minor axis direction but also along several other azimuthal angles.

\subsection{Expected effect of disk thickness to observed brightness}
\label{sec:expectedEffects}

The geometrical thickness of a protoplanetary disk can have different effects on its brightness, which we summarize in this sub-section. Fig.~\ref{fig:schematic} shows a schematic representation of the two main effects analyzed in this work. We consider here inclined disks or rings, with no intrinsic eccentricity and asymmetry.

\paragraph{Gap projection effect $-$ gaps and outer disk edges} While, after deprojection from the inclination, a gap in an infinitely thin disk will have the same shape at all azimuths, the depth and apparent width of a gap is expected to vary with azimuth in the case of a geometrically thick disk. If the disk is sufficiently inclined and vertically thick, a gap will appear less deep and less wide along the projected  minor axis than along the major axis direction (see middle \& right panels of Fig.~\ref{fig:schematic}). In other words, the depth of a gap in the projected minor axis cuts of an image can be dominated by the disk vertical thickness in inclined systems. This is the effect that has been previously analyzed for example by \citet{Pinte_2016}, \citet{Villenave_2022}, and \citet{Pizzati_2023}. It is most easily identifiable by using intensity cuts across the gap for different azimuthal angles. 
High angular resolution is critical for effectively resolving  the substructures and identifying those effects. For example, a typical angular resolution of 0.1'' corresponds to 15\,au at the distance of nearby star-forming regions (e.g., Taurus). At this resolution, most disk substructures remain unresolved, as is evidenced by the high fraction of smooth disks (20/32) found in the unbiased Taurus survey by \citet{Long_2018}. This underscores the need for higher angular resolution observations, as analyzed in this work (Appendix~\ref{app:observations}). 

A narrow gap configuration is the most favorable situation to identify this effect as the rings both interior and exterior to the gap can  be projected on to the gap and affect its width along the minor axis direction.
However, the vertical projection effect of a ring can also be visible at the outer edge of a disk. In that case, only the thickness of the disk interior to the outer (infinite) gap plays a role. In other words, the disk thickness is projected further out than the true (midplane) outer radius. A thinner disk will have a steeper outer edge profile along the minor axis direction than a thicker disk, and end at a smaller apparent radius. We refer to these effects as "G" in the case of the gap effect and "O" for the outer edge effect in \autoref{tab:results}.

\paragraph{Ring effect} 
For optically thin dust, a vertically thick, radially narrow ring is expected to exhibit azimuthal brightness variations. 
Although it may superficially look similar, this effect is different from the well-known limb darkening in stars, which requires optically thick emission and a temperature gradient. 
This mechanism has been discussed in detail for disks by \cite{Doi_Kataoka_2021}  
and relies on the amount of material which is crossed along the line of sight and which varies at different azimuths. 
Along the major axis the line of sight intersects both the width, $\sigma_r$, and the height, $\sigma_z$, of the ring. 
Along the minor axis, however, the projected material column density is reduced due to the projection of the vertical extent of the disk, resulting in lower observed intensities, such that 
$I_\text{minor}/I_\text{major} = \sigma_r / \sqrt{\sigma_r^2 + \sigma_z^2\tan^2 i}$~\citep{Doi_Kataoka_2021}. As a result, if dust is optically thin, the ring is predicted to be brighter along its major axis than along its minor axis (see left \& middle panels of Fig.~\ref{fig:schematic}). This effect is indicated by the letter "R" in \autoref{tab:results}.

However, we note that an elongated beam might artificially create a similar azimuthal variation in a ring's brightness  in observations. To confirm that the variations are due to a vertically thick ring rather than to an imaging artifact, radiative transfer models are required and synthetic images need to be created with similar characteristics as the observations. Alternatively, creating images where the beam is circular after deprojection from the disk inclination could alleviate this problem. 

\paragraph{Wall effect} When dust is optically thick and in the presence of a gap, the far side of the disk (wall), will be directly exposed to the observer. 
On the near side, the directly illuminated wall is hidden, leading to a crescent asymmetry along one side of the minor axis. The best example of disk walls are found in optically thick rings observed at scattered light wavelengths~\citep[e.g., GG Tau,][]{Duchene_2004}. At millimeter wavelengths, this effect has been clearly demonstrated  by \citet{Ribas_2024} in two systems, and also identified in other mid- or highly inclined disks \citep[e.g.,][]{Ohashi_2022, Lin_2023, Guerra-Alvarado_2024}. In mid-inclination systems it only becomes relevant close to the star ($\lesssim10$\,au).
This effect will thus not be specifically investigated in this work.

\subsection{Disk structure}
\label{sec:disk_structure}
We assume that the disk structure is axisymmetric. 
For all disks, we consider an inner radius of $R_\text{in} = 0.1$\,au and a fixed outer radius, $R_\text{out}$, as is indicated in \autoref{tab:disk_parameters}. The vertical density distribution of the gas is parameterized by a power law such that:
\begin{equation}
H_\text{g, model}(r) = 10~(r/100\,\text{au})^{1.1}\text{, for }R_\text{in}<r<R_\text{out}.
\label{eq:Hg}
\end{equation}
The dust is composed of astronomical silicates and graphite grains~\citep[same composition as in][]{Grafe_2013, Villenave_2023}. The distribution of dust sizes is fixed. Grains sizes $a$ are between 0.005\,$\mu$m and 3\,mm and follow a power-law distribution such that $n(a)\text{d}a\propto a^{-3.5}\text{d}a$. We consider 100 grain size bins with a logarithmic distribution in $a$. The gas-to-dust ratio is constant within the disk and assumed to be 100. 

We implement differential dust settling, using the prescription of \citet{Fromang_Nelson_2009}. This prescription describes turbulence as a diffusive process, and the vertical density distribution for a grain of size $a$ follows:
\begin{equation}
    \rho_\text{d}(r, z, a) \propto \Sigma_\text{d}(r)\exp\left[-\Sc\frac{\St_\text{MCFOST}(a)}{\alpha_{z, \text{MCFOST}}}(e^{\frac{z^2}{2H_\text{g}^2}} -1) -\frac{z^2 } {2H_\text{g}^2}\right],
    \label{eq:fromang}
\end{equation} 
where $\Sigma_\text{d}(r)$ is the dust surface density, and $H_\text{g}$ is the gas scale height (\autoref{eq:Hg}). 
$\St_\text{MCFOST}(a)$ is the Stokes number of the particles at the disk midplane ($z=0$): 
\begin{equation}
    \St_\text{MCFOST}(a) = \frac{\rho_\text{bulk} a}{\rho_\text{g, mid}H_\text{g}(r) }.
\label{eq:stokes}
\end{equation} 
The dust material density and gas density in the midplane are respectively described as  $\rho_\text{bulk}$ and $\rho_\text{g, mid}$. 
$\Sc$ is the Schmidt number, which quantifies the ratio of gas and dust diffusivities, and is fixed to~1.5~\citep{Pinte_2016, Fromang_Nelson_2009}. 
The degree of settling is set by varying the turbulence parameter $\alpha_{z, \text{MCFOST}}$.
For each disk, we consider 4 logarithmic levels of settling, with $\alpha_{z, \text{MCFOST}}$ between $2\times 10^{-4}$ and 0.2. Because this value is model-dependent, we later refer to the models as "Flat", "Thin", "Moderate", or "Thick",  respectively for $\alpha_{z, \text{MCFOST}}=2\times10^{-4}$,  $2\times10^{-3}$, $2\times10^{-2}$,  or $2\times10^{-1}$.

\subsection{Fitting the dust surface brightness}  
\label{sec:fittingSurfaceDensity}

We use the radiative transfer code \texttt{mcfost}~\citep{Pinte_2006, Pinte_2009} to model the brightness of the disks. For each disk, we set global (distance, sky coordinates), disk ($i$, PA, $R_{out}$), and stellar ($T_\text{eff}$, $R_\star$, $M_\star$) parameters based on observational constraints (\autoref{tab:disk_parameters}). 

We produced radiative transfer models of the disks based on an estimate of their surface density profile. To do so, we first estimate an initial surface density profile, as is described below. Then, we follow the approach of \citet{Pinte_2016} and \citet{Pizzati_2023}, and employed an iterative procedure to find a  surface density profile, allowing us to reproduce the  brightness profile along the major axis for each disk. 

We assumed that vertical projection effects are absent or limited in the major axis direction and used this to estimate the surface density of the disk. 
We extracted the major axis profiles  of the data $I(r)$ from their deprojected images, using a wedge with an opening angle of 30$^\circ$, averaged on both sides (i.e., for positive and negative radii). For HD143006, HD163296, and PDS70, however, we do not average both sides of the major axis profiles but instead  consider only the side without the crescent. 

An initial surface density profile was estimated using the following equation:
\begin{equation}
    \Sigma_\text{d}(r) = \frac{I(r)}{B_\nu[T(r)]\times \kappa_\nu},
    \label{eq:initialsurfacedensity}
\end{equation}
where $B_\nu$ represents the Planck function at a given temperature. To compute the initial surface density profile using \autoref{eq:initialsurfacedensity}, we assume a dust opacity of $\kappa_\nu = 13.27$ cm$^{2}$/g, independently of the observing wavelength, and an initial disk midplane temperature of $T=20 (r/100\,\text{au})^{-q}$, where $q =0.5$ for $r>5$\,au and $q=0.1$ for $r<5$\,au. These assumptions are only needed for the first iteration since the temperature structure and dust opacities are consistently calculated during the radiative transfer for all subsequent iterations.

Using this initial surface density, we generated a radiative transfer image with \texttt{mcfost}, for the same inclination and position angle as the data. Using these radiative transfer model images, we then produce synthetic images with the same uv-coverage as the data. This is done within CASA using the \texttt{ft} function. Then, synthetic model images are created using the \texttt{tclean} CASA task, with the same imaging parameters as the data. Finally, similarly to the data, we deproject the model image based on the disk inclination.

After extracting the major axis brightness profile of the deprojected model image, we produce a corrected dust surface density profile by multiplying the initial surface density profile by the ratio data over model. 
During that step, the total dust disk mass is also adjusted to match the absolute brightness of the major axis profile. We repeat the steps until the model brightness profile agrees with the data within 5\% or after 20 iterations, whichever is first.
We  report the final dust mass for each source in \autoref{tab:results}.   The reported value (resp. uncertainty) is the averaged (resp. standard deviation) dust mass over the four models with different thicknesses.

This method converges most effectively for  optically thin dust. For optically thick dust, the method becomes ambiguous as iterations can increase the dust mass without a corresponding increase in observed flux, and the models can fail to reproduce the major axis brightness.
This is why, although included in the model, the iterative procedure is only applied between 5\,au and $R_\text{out}$ to avoid modeling the optically thick inner region. 
Moreover, for a few systems (GO Tau, GW Lup, HL Tau, Oph163131), the stellar parameters inferred by the literature do not allow the models to be bright enough to reach the observed flux in some bright dust rings. This is because with the literature stellar parameters, the dust temperature is too low for the models to reproduce the observations. 
To allow the models to converge, we decided to increase the disk temperature in order to have sufficiently bright dust in the outer disk. 
We performed this by artificially increasing the stellar radius, and thus the stellar luminosity, in the modeling (systems marked with $^*$ in column $R_\star$ of \autoref{tab:disk_parameters}).
Physically, the stellar luminosities may have been under-estimated in the case of the presence of an envelope (e.g., HL Tau) or an edge-on disk (e.g., Oph163131). Alternatively, a different disk density profiles, not explored in this work, may allow for a warmer midplane for similar stellar parameters. The final models have been uploaded to a zenodo repository  (\href{https://doi.org/10.5281/zenodo.14056832}{zenodo.14056832}).

\subsection{Identifying the best model utilizing different azimuthal wedges and azimuthal profiles}
\label{sec:bestmodel_def}

Using the iterative procedure previously described (Sect.~\ref{sec:fittingSurfaceDensity}), we generated four models per disk, with different levels of dust sedimentation (parametrized by $\alpha_{z, \text{MCFOST}}$). 
All of them match the major axis profiles of the observations, while having different vertical dust height. To identify the effect of the dust vertical height on the global disk appearance, we compared data and model profiles along different azimuths.  From the deprojected data and model maps, we obtained six wedges of 30$^\circ$. They go from the minor axis to 15$^\circ$ from the major axis, allowing us to investigate different levels of projection effects along several directions. 

To identify the best fitting model, or the models to be excluded, we first performed a visual check of the different cuts. We focused on specific regions in the disks. Specifically, we defined regions centered around the gaps or including the outer disk, to look for the expected projection effects (Sect.~\ref{sec:expectedEffects}). These regions of interest were defined visually based on the major axis profile and will allow us to study if there is any variation in the settling efficiency within a given disk. 

Following our initial visual identification of the regions of interest and subsequent identification of inappropriate models, we employed two quantitative metrics to assess the discrepancies between the models and the data.
The first metric consists of a normalized $\chi^2$ parameter and the second criterion, $\mathscr{S}$, quantifies the difference in shapes between the data and model curves. For the first metric, we average $\chi^2_\text{azimuth} / \chi^2_\text{majorAxis}$ over each azimuthal direction. This quantifies the distance of the model with the data but does not consider the difference in shape. The second metric estimates the Hausdorff distance \citep[$\mathscr{S}$,][]{munkres_topology_2000}, which takes into account the likelihood in shape between the model and the data. This metric measures the "distance" between two sets of point by identifying the maximum distance between any point in one set and its nearest neighbor in the other set. 

By visual inspection, we identify that models with $\chi^2 \leq 8$ and $\mathscr{S} \leq 1.2$ are generally good models. In addition, we also consider the shape of the $\chi^2$ and $\mathscr{S}$ curves as a function of $\alpha_{z, MCFOST}$ to identify trends for better/worse models with thickness. The constraints reported in \autoref{tab:results} reflect the combined use of the metrics and of our visual analysis to exclude models. In some systems, the $\chi^2$ or the $\mathscr{S}$ metric is not discriminating and is thus not considered. Appendix~\ref{app:cuts} shows the final values of $\chi^2$ and $\mathscr{S}$ for each model, as well as the major axis profiles and other relevant  azimuthal profiles providing constraints on the dust scale height, for all disks.

Finally, we note that for three disks discussed in Sect.~\ref{sec:thickthin}, we also generated azimuthal profiles along their rings, to investigate the ring effect presented in Sect.~\ref{sec:expectedEffects}. For the inner ring of these disks, the constraint to the vertical thickness was mainly obtained by visual inspection.

\subsection{Getting the dust scale height for the best model}
\label{sec:relevantquantities} 

 Once the best model was identified, we estimated its dust scale height, $H_\text{d}$.  We consider only the grains that emit the most at the wavelength of the observations (Appendix~\ref{app:observations}). Those grains are identified within \texttt{mcfost}, based on the dust properties assumed in the modeling, and have typical sizes around 100\,$\mu$m (\autoref{tab:results}).
 To estimate their dust scale height,  we fit the \texttt{mcfost} vertical distribution of these grains, at each radius, with a 1D Gaussian. This provides us with a radial profile of the relevant dust scale height. Then,  we obtained the average dust scale height in each radial region of interest, which we divide by the average radius of the region to obtain the aspect ratio $H_\text{d}/R$. 

For completeness, we also report the corresponding  $\alpha_{z, \text{MCFOST}}$ of the constraining model and the Stokes number of these particles assumed in the modeling, $\St_\text{MCFOST}$, in \autoref{tab:alpha}.  However, we emphasize that the dust vertical thickness is the only quantity that can be directly constrained by the observations. If different assumptions for the Stokes number were considered in the modeling (for example because of a different dust surface density or gas-to-dust ratio), $\alpha_{z, \text{MCFOST}}$ would change, but the geometrical effects mentioned in Section~\ref{sec:expectedEffects} would still be observed for the same dust height.

\section{Modeling results}
\label{sec:results}

Our modeling allows us to constrain the dust vertical extent in 23 out of the 33 systems considered. We identify five disks with both a lower limit on the dust scale height at some inner radii and an upper limit on their dust height in the outer regions, indicative (for three of them) of variation in thickness with radius (see Sect.~\ref{sec:thickthin}). In the 18 remaining systems, our modeling provides  useful upper limits on the dust scale height in the outer disks (see Sect.~\ref{sec:upperlimits}). 
We summarize the constraints obtained on the dust scale height from these 23 disks in \autoref{tab:results} and Fig.~\ref{fig:alpha_constraints}. Additional figures representing the results are shown in Appendix~\ref{app:cuts}. We compare the results with previous studies in Sect.~\ref{sec:previousstudies}.

\begin{figure}
    \centering 
    \includegraphics[width = 0.47\textwidth]{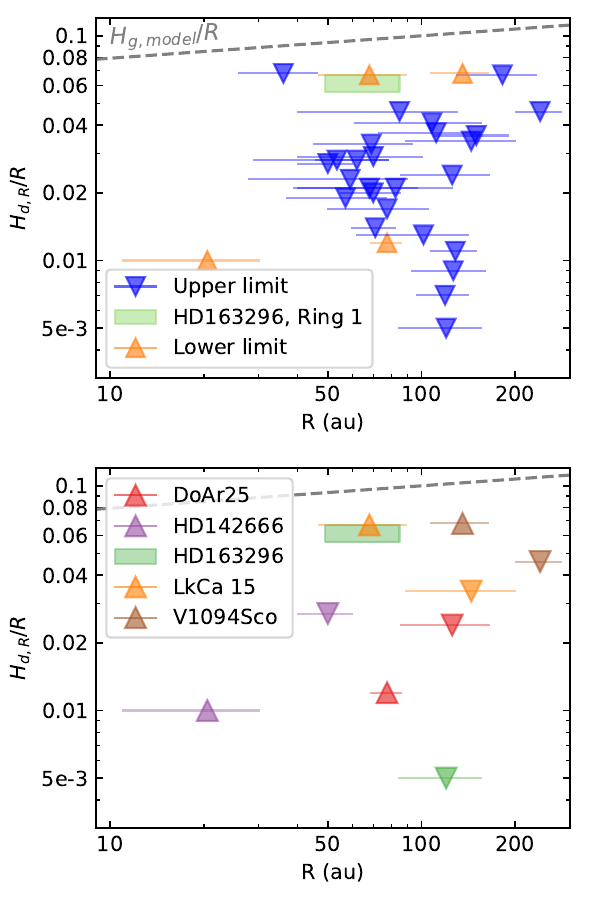}
    \caption{Constraints on the vertical concentration of dust as a function of radius. Each symbol is associated with an horizontal bar corresponding to the range of radii over which the constraint was obtained. The gas scale height assumed in the modeling (\autoref{eq:Hg}) is indicated as a dashed gray line. \emph{Top:} All disks. \emph{Bottom:} Five systems with both lower and upper limits on their dust height. }
    \label{fig:alpha_constraints}
\end{figure}

\subsection{Finite dust height and variation in thickness with radius in 5 disks}
\label{sec:thickthin} 

\begin{figure*}
    \centering
    \includegraphics[width = 1\textwidth, trim={0cm 0.9cm 0 0},clip]{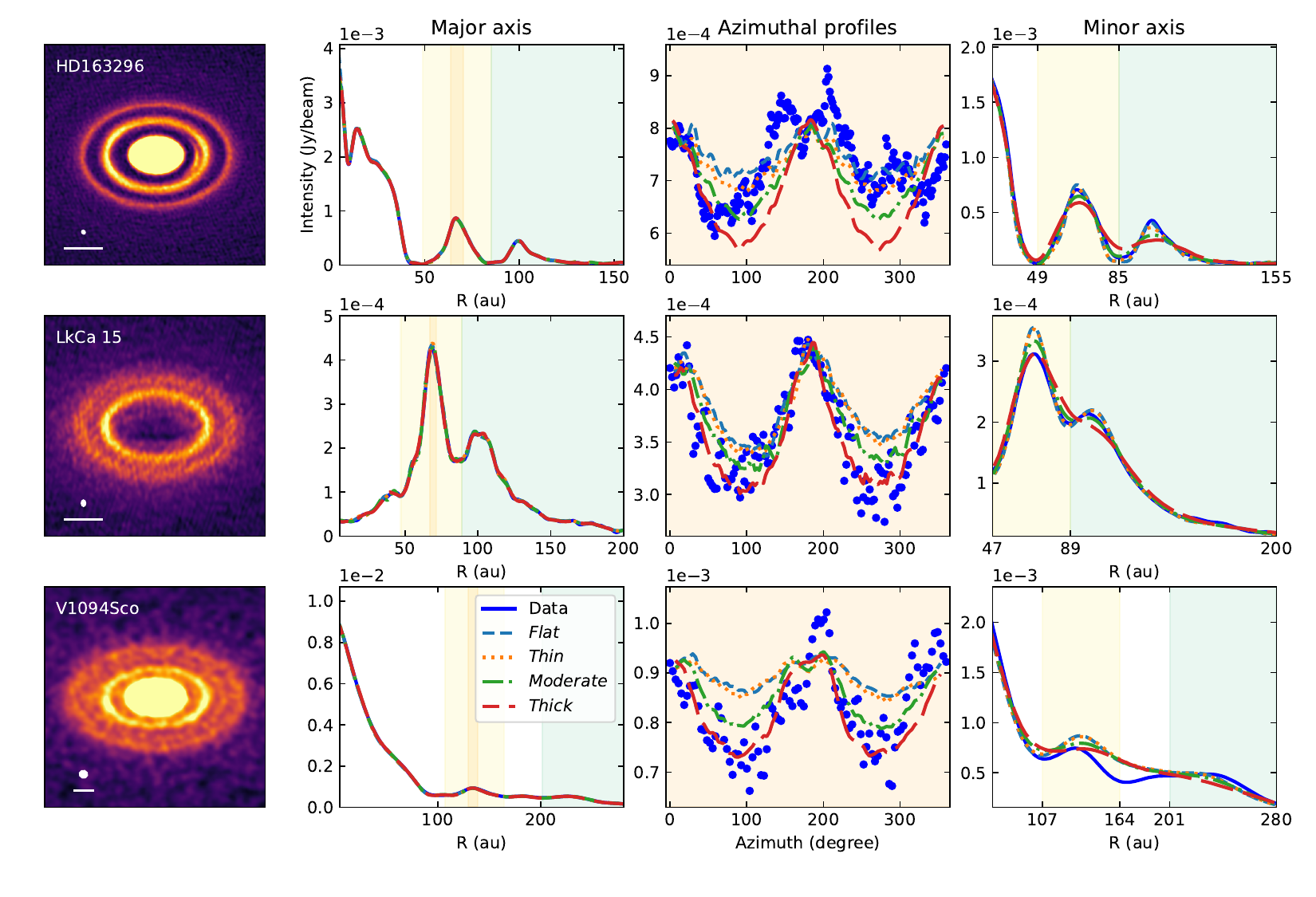}
    \caption{Vertically thick inner ring and thin outer disk in HD163296, LkCa 15, and V1094Sco. The shaded colors on the major axis cuts indicate the location of the azimuthal profiles (orange) and rings of interest (yellow and green).  
    "Flat" models are too thin vertically to show sufficient azimuthal variation compared to the data (third panels), while "Thick" models  are too vertically extended to reproduce the outer disk/ring brightness in all three disks. The beam size and a 25\,au scale are indicated in the bottom left corner of the first panels.}
    \label{fig:upperLower_LkCa15_V1094Sco}
\end{figure*} 

\begin{figure*}
    \centering
    \includegraphics[width = 1\textwidth, trim={0cm 0.4cm 0 0},clip]{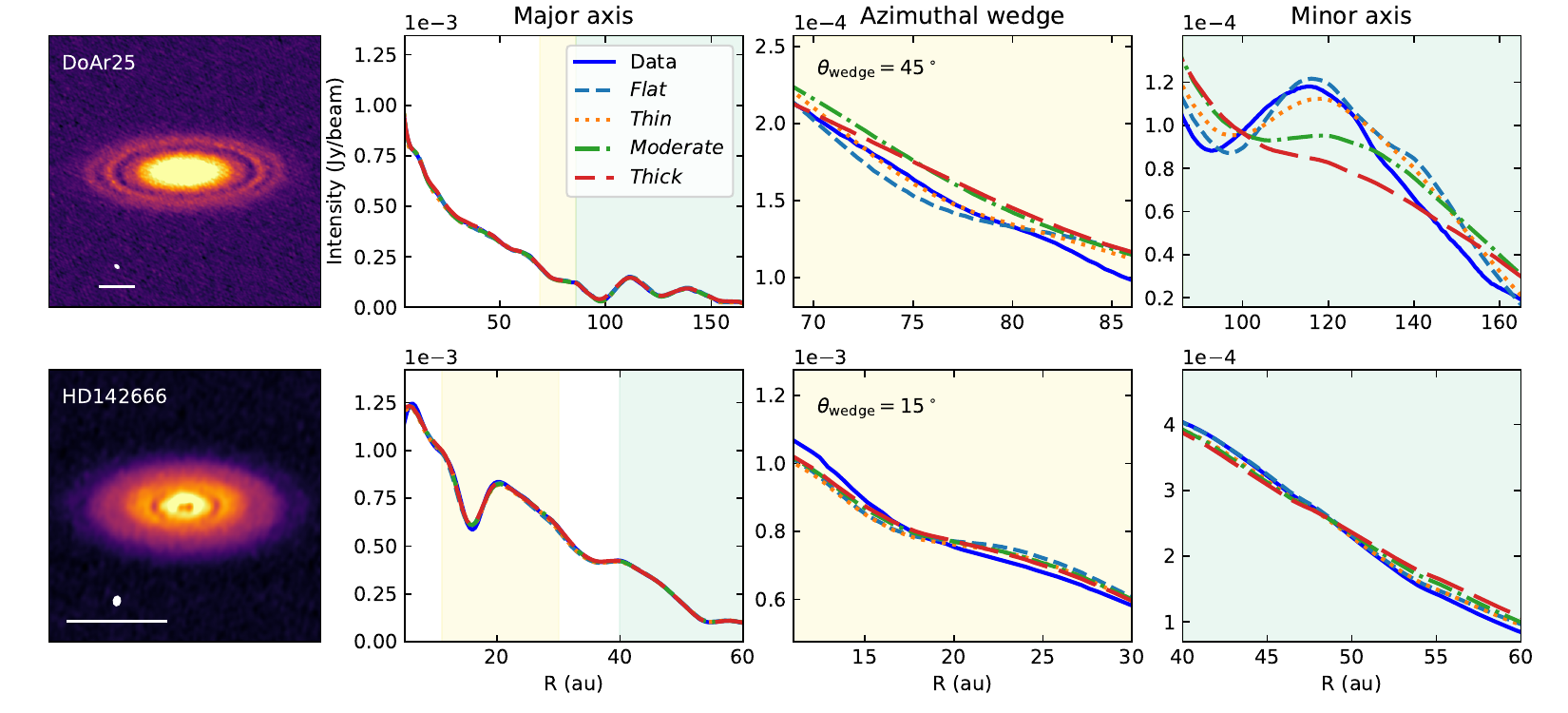}
    \caption{Resolved thickness in the inner disk half and flat outer disk in DoAr~25 and HD~142666. For both disks, "Flat" models show too strong of a gap in the inner region (third panels), indicating that the disks must be thicker. On the other hand, "Moderate" and "Thick" models do not reproduce the gap and ring in DoAr~25, and the outer disk profile in both disks (last panels), highlighting the presence of a flat outer disk. The beam size and a 25\,au scale are indicated in the bottom left corner of the first panels.}
    \label{fig:upperLower_DoAr25_HD142666}
\end{figure*}

We identify five disks for which the scale height estimates vary with the orbital radius: DoAr~25, HD~142666, HD~163296, LkCa~15, and V1094~Sco. For all of them, the region typically within the inner half of the disk shows a finite vertical height, while an upper limit on the scale height is obtained for their outer disks. 
Here, "finite vertical height" refers to regions where we have either measured the dust scale height or established a lower limit. These regions may still exhibit some settling, but they contrast with other regions where we can only place upper limits on the dust thickness.  
We describe below how the constraints were obtained for each system (see also Fig.~\ref{fig:upperLower_LkCa15_V1094Sco}, Fig.~\ref{fig:upperLower_DoAr25_HD142666}, and Fig.~\ref{fig:finalCuts_DoAr25_HD142666_HD163296_LkCa_15_V1094Sco}).  

The constraints in the inner ring of HD~163296, LkCa~15, and V1094~Sco were obtained using the ring projection effect. Similarly to previous results on HD~163296~\citep{Doi_Kataoka_2021, Doi_Kataoka_2023, Liu_2022}, we identify significant azimuthal variations in the inner or intermediate ring brightness. The azimuthal profiles along their respective rings (at 49--85\,au, 47--89\,au, and 107--164\,au, respectively) are shown in the second right panel of Fig~\ref{fig:upperLower_LkCa15_V1094Sco}. They had not been previously reported for LkCa 15 and V1094Sco. In all three cases, the data display significant azimuthal variations in the ring brightness, with a maximum along the major axis and a minimum along the minor axis.  
As is presented in Sect.~\ref{sec:expectedEffects} and Fig.~\ref{fig:schematic}, this is expected in the case of a vertically thick (optically thin) ring. 
Indeed, we find that this azimuthal variation is not reproduced by the thinnest models ("Flat"), while thicker models correspond better to the data.

On the other hand, in the case of DoAr~25 and HD~142666, we used geometrical projection effects in the innermost gap (at 69--89\,au, 11--30\,au, respectively) to show that the thinnest models are too thin to reproduce the data. 
In both systems, the inclination is  large ($>60^\circ$) and the inner gap is relatively shallow so the best constraints were obtained along an azimuthal direction different from the minor axis direction. The results are displayed in Fig.~\ref{fig:upperLower_DoAr25_HD142666}, with a zoom in the inner disk region along a constraining azimuth in the second-right panels. In both DoAr~25 and HD~142666, the gap is mostly undetected in the data along azimuths of 45$^\circ$ or 15$^\circ$ away from the minor axis direction. However, the thinnest model ("Flat") 
still shows the gap clearly along these azimuths indicating that the disk must be thicker in this inner region. This allows us to exclude this model. However, none of the thicker models presents a clear best match, allowing us to only place a lower limit on the allowed dust scale height in the inner region of the disks to $H_{\text{d}, R}/R \geq0.01$.

In contrast, for all five systems, the outer disks are best reproduced by the thinner models (see Fig.~\ref{fig:upperLower_LkCa15_V1094Sco} and Fig.~\ref{fig:upperLower_DoAr25_HD142666}). 
In DoAr25, the gaps and ring between 86 and 135\,au are not visible along the minor axis direction for "Moderate" and "Thick" models, implying that the outer disk must be vertically thin. 
Considering the outer disk edge or the ring effect, we find that the outer regions of the other disks are also significantly vertically thin, with an upper limit of $H_\text{d,R}/R \leq 0.046$  in all cases, but as low as $H_\text{d,R}/R \leq 0.005$ in the outer ring of HD163296 (\autoref{tab:results}). The dependence of the dust height with radius  ($H_\text{d, R}/R$) in these systems is displayed in the bottom panel of Fig.~\ref{fig:alpha_constraints}. 

To summarize, we identify five disks with both a finite thickness in their inner disk ring and a thin outer disk.  For three out of these five disks (HD163296, LkCa 15, and V1094Sco), the relative aspect ratio, $H_{\text{d}, R}/R$, in the inner ring is larger than that in the outer ring, while this is not necessary the case for the other two systems. 
For two disks, an inner gap is used to constrain the thickness of the inner disk, while azimuthal variation in the inner ring brightness is used in the other three systems. The outer regions of these five systems are found to be significantly affected by vertical settling with levels similar to the other disks studied in this work~(Sect.~\ref{sec:upperlimits}).

\subsection{Vertically thin outer disks in 18 systems}
\label{sec:upperlimits}

Similarly to previous studies, most systems included here only allow us to place upper limits on the acceptable value of the dust disk height. In the majority of cases, the best models correspond to those with the thinnest vertical extent (i.e., the weakest vertical turbulence $\alpha_{z, \text{MCFOST}}$). 
We show the major and minor axis profiles of the data and models in the regions where we obtained some constraints in Appendix~\ref{app:cuts}. The results are also summarized in \autoref{tab:results}. In most cases, the upper limits on the dust height were obtained using the gap projection effect (see Sect.~\ref{sec:expectedEffects}) either in a gap or in the outer edge of the disks.

Out of the 18 disks with only upper limits on their vertical height, only 3 allowed us to obtain different upper limits between the inner and outer regions of the disks. In AA Tau and AS209, the dust scale height is constrained to a smaller value  in the outer regions than in the inner regions. On the other hand, the constraint in the outer ring of CI Tau is less strong than in its inner regions. In summary, no clear trend is obtained between the ring location and the upper limit on the dust scale height obtained for these systems (Fig.~\ref{fig:alpha_constraints}).

\subsection{No constraints for 10 disks}

While we selected disks with morphologies and orientations expected to be favorable for the study their vertical structures, a subsample of them do not allow us to place such constraints. Those are eight DSHARP targets (DoAr~33, Elias 20, Elias~27, HD~143006, SR~4, Sz~114, WaOph~6, WSB~52), one source from the Taurus survey (MWC~480), and DM Tau. These sources typically present lower inclinations and/or shallower gaps than in the rest of the sample, hence minimizing projection effects and preventing us from extracting information on their vertical structures. Additionally, we are unable to constrain the vertical distribution of the spiral disk Elias~27 because the gap interior to the spirals becomes deeper along the minor axis, which our radiative transfer model is unable to reproduce. This is likely caused by a true azimuthal variation in the dust density in this gap. 

\subsection{Comparison with previous millimeter studies}
\label{sec:previousstudies}

The analysis presented in this work is  new for 17 of the 33 disks of the sample. We extract new constraints on the dust vertical height for 13 of these 17 systems. The other ten disks for which we constrain the dust height have been previously modeled. Here, we compare our results with literature results for these ten disks. 

Our results are consistent with the constraints obtained by \cite{Pinte_2016} on HL Tau, who  used the same prescription for vertical settling and devised the modeling approach used in the present work. 
However, the comparison of their results and our \autoref{tab:results} is not direct, because \cite{Pinte_2016} report the height of 1mm particles at a radius of 100au ($H_\text{1mm}=0.7$\,au), while we report the height of the grains emitting the most at the observing wavelength, that is $a\sim59\,\mu$m in the case of HL Tau, at the location of the constraining region. 
Our constraints translate to $H_\text{1mm}\leq1.9$\,au at 100au, which is in agreement with the results by \cite{Pinte_2016}, although  less constraining. Another difference between the two works is also that they fitted observations at bands 3, 6, and 7 together, while we only consider band~7 observations in this analysis. Our analysis is therefore sensitive to (less settled) smaller grains and higher optical depth, which could lead to a higher dust scale height. 
We also note the discrepancy in the reported $\alpha_{z, \text{MCFOST}}$ of a factor about 7 between both studies. This is due to a difference in the total dust mass in both models, 
and allows us to highlight again that $\alpha_{z, \text{MCFOST}}$ is a model-dependent estimate of the turbulence.

Our modeling of AS~209, Elias~24, GW Lup, MY Lup and Oph~163131 are consistent with constraints obtained in previous studies~\citep{Pizzati_2023, Villenave_2022}. More specifically, we obtain similar upper limits for Elias~24 and the inner ring of AS~209 than in \citet{Pizzati_2023}. Our analysis of the second ring of AS209, GW Lup, and MY Lup however allows us to improve on the constraint limits placed by \cite{Pizzati_2023} by a factor 1.4 to 3.1. This is due to the fact that our analysis now takes into account not only the minor axis profile but also different azimuthal directions.  On the other hand, \citet{Villenave_2022} obtained a smaller upper limit on the dust scale height of Oph163131 by a factor of about 2 compared to our result. This is because they exploit the unique geometry of this nearly edge-on system, focusing on the minor axis cut through the `ansae' of the gap to estimate the vertical extent. In contrast, we employ a more general approach applicable to a wider range of systems. 

In the case of HD 163296, our results slightly differ from the literature results. We obtain both a lower and an upper limit on the dust vertical scale height in the inner ring ($0.006 \leq H_{\text{d}, R}/R\leq 0.067$), while  previous studies only constrained a lower limit on the scale height of this ring~\citep[$H_{\text{d}, R}/R\geq 0.056$,][]{Doi_Kataoka_2021, Liu_2022}. 
Because the lower limit obtained from previous studies is more constraining than our results, in the remaining of this study, we consider it as the lower limit on the dust scale height of HD 163296, such that:  $0.056 \leq H_{\text{d}, R}/R \leq 0.067$.

We note that six of the ten disks where our modeling does not allow to obtain constraints were included in the sample of \citet[][DSHARP disks without spirals]{Pizzati_2023}. This work was also unable to estimate their vertical height using the gap projection effect. They empirically concluded that disks with an inclination of $i>25^\circ$ and a gap depth lower than 0.65 are most favorable for constraining their vertical extent. \\

Previous studies at (sub)millimeter wavelengths already identified hints toward thicker inner disks in  some of the systems considered here. For example, interior to the innermost gap of HD~142666, the minor axis profile presents an asymmetric brightness (previously pointed out by \citealt{Jennings_2022} and \citealt{Ribas_2024}). These crescent-shaped asymmetries have been interpreted as being due to an optically thick wall~\citep[e.g,][Sect.~\ref{sec:expectedEffects}]{Ribas_2024}, which is consistent with our finding of a finite dust height around the inner gap of this system.

Interestingly, \cite{Guerra-Alvarado_2024} analyzed shorter wavelengths observations of HL Tau (band 9, 0.45\,mm) and also identified a crescent feature, located around 32\,au. 
They estimated that the scale height of the dust particles should be $H_{\text{d}, 32}/R> 0.08$, which is significantly larger than our constraints at radii greater than 61au ($H_{\text{d}, R}/R\leq0.041$). This suggests that HL Tau might also be geometrically thicker in the inner regions than in the outer regions. Alternatively, the difference could be due to the fact that band 9 probes smaller grains than our band~7 observations, which are more vertically extended  
in the disk and have a larger optical depth than those detected in band 7.   
Similarly, our constraints on RY~Tau at large radii (28--90\,au, $H_{\text{d}, R}/R \leq 0.046$) can be compared to the results of \cite{Ribas_2024} who analyzed the required height of dust at 12\,au to reproduce the wall effect observed in this system. Using radiative transfer modeling, they obtain $H_{\text{d}, 12} = 0.2$\,au at 12\,au, corresponding to $H_{\text{d}, 12}/R\sim 0.017$. This value is consistent with our upper limit at radii larger than 28\,au.

\subsection{Comparison with scattered light observations} 

In this work, we find that five disks present a finite dust height in their inner region (i.e., measured or lower limit on the dust scale height) and an unresolved height in the outer disk. 
Three of them (HD~163296, LkCa 15, and V1094~Sco) definitively show that the inner ring is thicker than the outer region. Such spatial variation in the disk thickness with radius, if also present for smaller particles, is expected to leave signatures on their scattered light images. Indeed, a thick inner region would cast a shadow to the outer disk, making it undetected at scattered light wavelengths.

Interestingly, these five disks have been observed with the VLT/SPHERE in scattered light. All disks are well detected and bright at near-infrared wavelengths. DoAr~25, HD~142666, and V1094~Sco show scattered light sizes comparable or larger than the millimeter dust emission, without clear substructures~\citep{Garufi_2020, Garufi_2022} excluding disk self-shadowing. On the other hand, HD~163296 and LkCa~15 show a morphology clearly distinct from that of the continuum dust emission~\citep{Muro-Arena_2018, Thalmann_2016}. In both cases, only the innermost ring is detected in scattered light, consistent with self-shadowing from the inner ring.  
\cite{Muro-Arena_2018} modeled both the ALMA and SPHERE images of HD~163296. They showed that increased settling of the dust grains or depletion of small grains in the outer ring can explain the absence of outer ring in the scattered light images. In both HD~163296 and LkCa~15, the appearance of the scattered light images is thus suggestive of a thicker inner ring casting a shadow to the outer disk.
Even though these studies consider very different grain sizes, this geometry is similar with what we inferred from the analysis of the 1.25\,mm observations of these two systems. Detailed multi-wavelength modeling of these two systems could allow to further explore the physical relationship between the disk appearance in both wavelength ranges.  

We also checked the ratio of near-infrared excess (NIR) to far-infrared excesses  (FIR) for disks in our sample, as it could be indicative of some level of self-shadowing. Specifically, FIR-excesses less than 10\% coupled with NIR excesses above 10\% have been interpreted as evidence for self-shadowing~\citep{Garufi_2024}. We find that 16 disks in our sample have published estimates for both the near and the far infrared excess. Seven of them fall into the category of potentially self-shadowed systems~\citep{Garufi_2018, Garufi_2022, Garufi_2024}. Those are CI Tau, GW~Lup, WSB52, DS Tau, and 3 of the sources where we identified a finite dust height in their inner regions at millimeter wavelengths: HD142666, HD163296, and LkCa 15. Besides, both DoAr25 and V1094Sco are bright in the near infrared and faint in the far infrared~\citep{Garufi_2020}, which would be consistent with the presence of a thicker inner region as found in our study.

\section{Constraints on anisotropy of turbulence}
\label{sec:constraintsturbulence}

\subsection{Vertical dust-gas coupling}
\label{sec:vertical_turbulence}

In Sect.~\ref{sec:results}, we showed that most outer disks are vertically thin, which suggests that vertical settling is occurring. Here, we aim to constrain the level of settling in the different regions. We estimate the ratio $\alpha_z/\St$, which involves the level of vertical turbulence $\alpha_z$, and the Stokes number of the particles $\St$ characterizing the dust--gas coupling. Particles which are well mixed with the gas have a large ratio, while a low ratio ($\alpha_z/\St<1$) implies strong decoupling and significant vertical concentration. 
Assuming a balance between vertical stirring and settling, the ratio $\alpha_z/\St$ relates to the dust and the gas scale height, respectively $H_\text{d}$ and  $H_\text{g}$, following~\citep[e.g.,][]{Dubrulle_1995}:
\begin{equation}
    \frac{\alpha_z}{\St} = \left[ \left(\frac{H_\text{g}}{H_\text{d}}\right)^2 -1 \right]^{-1}.
\end{equation}

To estimate this ratio for all disks, we first estimate the gas scale height at the different ring locations. We report two scale heights and $\alpha_z/\St$ ratios in \autoref{tab:verticalTrapping} based on different assumptions, as is described below. On the one hand, we estimate  $H_\text{g, model}$, the gas scale height as assumed in the modeling, which is using \autoref{eq:Hg}. The corresponding $\alpha_z/\St|_{model}$ ratio was obtained using this gas scale height.

On the other hand, we also estimate $H_\text{g, temp}$, the gas scale height obtained from the hydrostatic equilibrium using the midplane temperature of the models, $T$, following:
\begin{equation}
    H_\text{g, temp}(r) = \sqrt{\frac{k_\text{B} Tr^3}{ GM_\star\mu m_\text{p}}},
    \label{eq:Hg_temp}
\end{equation}
with $k_\text{B}$ the Boltzmann constant, G the gravitational constant, M$_\star$ the stellar mass, $\mu =2.3$ the reduced mass, and $m_\text{p}$  the proton mass. 
To obtain the values reported in the second to last column in \autoref{tab:verticalTrapping}, we consider the temperature of the most constraining model, averaged over the radial domain of the constraint. We find that the gas scale height obtained based on the hydrostatic equilibrium is systematically lower than the scale height assumed in the modeling, by up to a factor 2 lower. In addition, for each disk, we checked that the adopted $H_\text{g, temp}$ agrees within 10\% with the average scale height obtained using the temperature of the 4 models with different vertical thicknesses.   We report the corresponding $\alpha_z/\St|_\text{temp}$ ratio in \autoref{tab:verticalTrapping}.  

\begin{figure*}
    \centering
    \includegraphics[width = 0.99\textwidth]{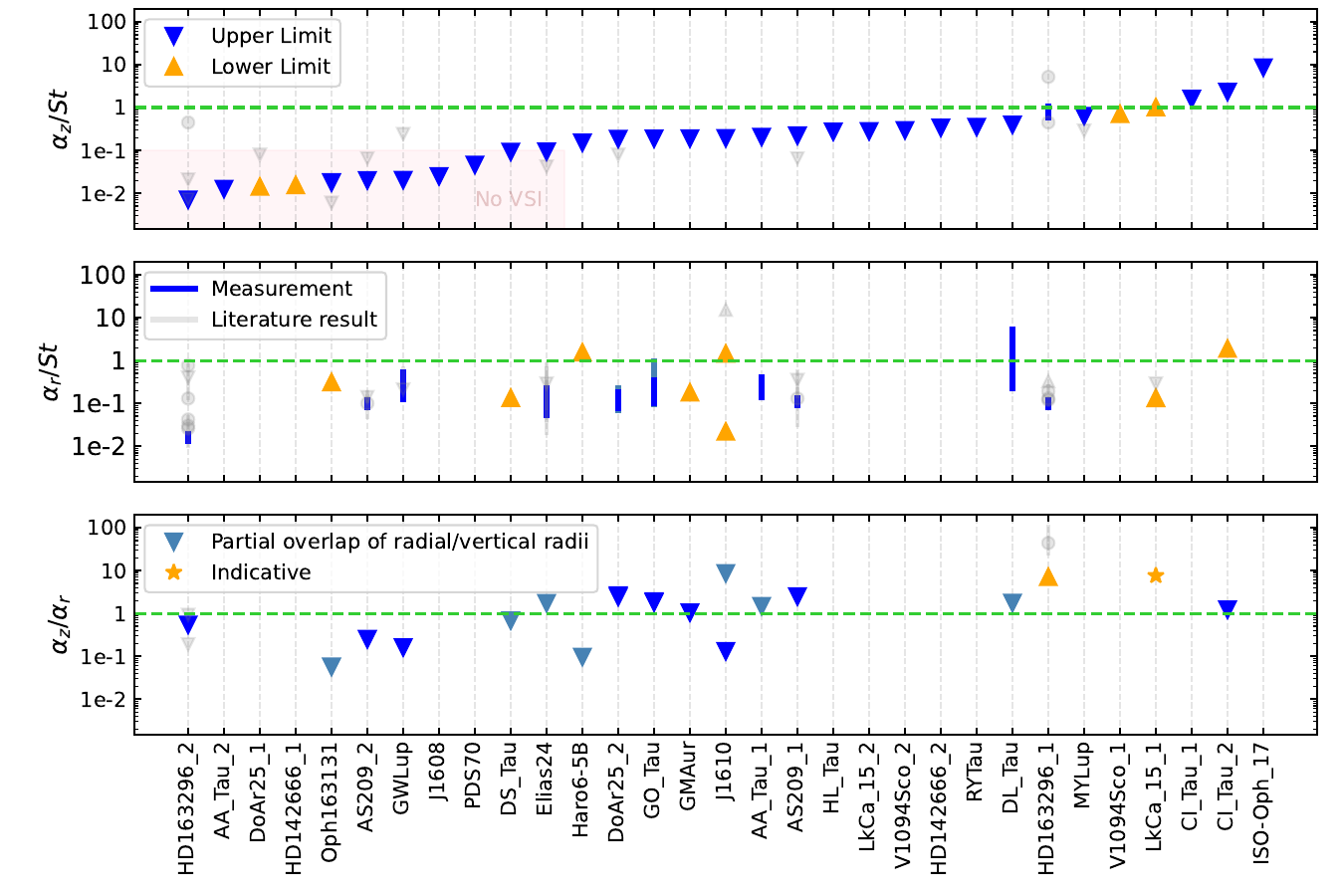}
    \caption{Vertical $\alpha_z/\St$ (top), radial $\alpha_r/\St$ (middle), and their ratio $\alpha_z/\alpha_r$ (bottom). In the bottom panel, we mark the rings without a full inclusion into the domain of the vertical constraints with a lighter blue. In all panels, constraints obtained by previous studies are shown in gray. The labels correspond to the constrained regions in \autoref{tab:verticalTrapping}. The underscore `\_1' corresponds to the only or first region with a constraint in a disk, while `\_2' displays the results for the second region.} 
    \label{fig:radial_vertical_alpha}
\end{figure*}

The top panel of Fig.~\ref{fig:radial_vertical_alpha} shows the most conservative values for the vertical $\alpha_z/\St$ in each region, ordered by increasing coupling. Nearly all disks/rings show $\alpha_z/\St<1$, which indicates that the grains are affected by vertical settling. Our results are consistent with those of previous studies~\citep[GW Lup, DoAr25, Elias24, AS209, MY Lup; HD164296; Oph163131;][]{Pizzati_2023, Doi_Kataoka_2023, Villenave_2022}. Interestingly, however, some of the thick rings (HD163296, V1094Sco, and LkCa~15) have lower limits on their $\alpha_z/\St$ ratio close to 1, indicating that settling is not efficient in these regions. \\ 

We then used these results to investigate potential correlations between the settling level, characterized by the ratio $\alpha_z/\St$, and various disk or stellar parameters. As has already been shown in Fig.~\ref{fig:alpha_constraints}, there is no clear correlation between the radius and the local dust scale height between systems. Moreover, although we do not show the figures here, we also find no clear correlations with parameters such as accretion rates, dust masses, stellar masses, and the dust to stellar mass ratio. 
Specifically, we see that all disks with a finite dust scale height are spread across a wide range of accretion rates, dust and stellar masses (between 0.84 and 2.06\,M$_\odot$). 
Finally, we also note that our study  includes 8 transition disks: AA Tau, GM Auriga, HD163296, J1608, J1610, LkCa 15, PDS70, and RY Tau.  Our sample, however, does not reveal significant differences between the level of vertical settling between the transition disks and the disks without a large central cavity. Similarly, the binary system Haro6-5B does not stand out from the rest of the sample, which is mostly composed of disks around single stars.  

\subsection{Radial dust--gas coupling}

In the radial direction, at the location of dust rings, the equilibrium between dust drift and turbulent spreading also allows to relate a radial $\alpha_r/\St$ to the dust and gas radial width, respectively $w_\text{d}$ and $w_\text{g}$~\citep[e.g.,][]{Dullemond_2018}:
\begin{equation}
    \frac{\alpha_r}{\St} = \left[ \left(\frac{w_\text{g}}{w_\text{d}}\right)^2 -1 \right]^{-1}.
\end{equation}

To estimate this ratio, we first evaluated the radial width of the dust rings. Previous studies aiming to constrain $w_\text{d}$ have either worked in the image plane~\citep{Dullemond_2018}, which is subject to uncertainties regarding the unconvolved width, or were performed using some visibility fitting~\citep[e.g.,][]{Jennings_2022}. Visibility fitting systematically reaches lower values than what was estimated from the image plane. Here, we took an alternative approach, where we consider the surface density profiles of the models as an unconvolved representation of the data. For each disk, we estimated the radial width of the rings by fitting a Gaussian to the rings in the final surface density profile obtained in the modeling (see Sect.~\ref{sec:model}). 

The results are reported in \autoref{tab:radialTrapping}. We note that not all the rings for which a constraint in the vertical direction exists could also be constrained in the radial direction. In many cases, this is because the surface density profiles do not resemble a Gaussian, or because multiple rings are present in the unconvolved model to reproduce one single ring of the observations. The uncertainties reported in \autoref{tab:radialTrapping} correspond to the standard deviation of the fit for all four models of each disk. \\

To estimate the $\alpha_r/\St$ ratio, we now need to estimate the radial width of the gas, $w_\text{g}$. This could in principle be done directly by using molecular gas observations of the disks and tracing for deviation from Keplerian velocity. We use such constraints for the two disks studied by \cite{Rosotti_2020}: AS209 and HD163296. 
However, such data are not always available for our sample and the complete analysis is out of the scope of this work. Instead, for the rest of the sample, we follow the simpler approach of \cite{Dullemond_2018} in order to obtain a reasonable upper and lower limit for the radial width of the gas. 

Under the assumption that the rings correspond to pressure bumps, the minimal size of the gas width has to be the gas scale height~\citep[e.g.,][]{Dullemond_2018}.  We consider the local gas scale height as the lower limit for the gas radial width. Here, we only take into account the hydrostatic equilibrium gas scale height obtained using the model's midplane temperature ($w_\text{g, min} = H_\text{g, temp}$) because it is the lowest (see \autoref{tab:verticalTrapping}) and thus provides the most conservative estimate for $\alpha_r/\St$. 
On the other hand, following \cite{Dullemond_2018}, we assume that the distance from the peak of the ring to the deepest point in the gap inside of the ring is the upper limit for the gas width ($w_\text{g, max}$). The results are reported in \autoref{tab:radialTrapping}.

We display the estimates for $\alpha_r/\St$ in the second panel of Fig.~\ref{fig:radial_vertical_alpha}. Those were obtained for 21 rings, located in 15 disks (\autoref{tab:radialTrapping}). We find that 11 rings are consistent with radial dust trapping ($\alpha_r/\St<1$), while the other 10 rings do not necessarily require strong trapping. We find a good agreement in the resulting constraints for the dust rings in common between our sample and previous studies~\citep[AS209, Elias24, GW Lup; HD163296; J1610;][]{Dullemond_2018, Rosotti_2020, Jennings_2022, Carvalho_2024, Doi_Kataoka_2023, Facchini_2020}. 

We are able to obtain a constraint on the radial trapping in only two vertically thick rings, in HD163296 and LkCa~15 (using $w_d$ value from \citealt{Facchini_2020}). In LkCa~15, the apparent dust width is larger than the gas scale height allowing us to only place a lower limit on the ratio $\alpha_r/\St$. On the other hand, similarly to results from \cite{Doi_Kataoka_2023}, we find that the inner ring of HD163296 is subject to some radial trapping. 

\subsection{Anisotropy of turbulence}
\label{sec:anisotropyTurbulence}

We used our constraints on the vertical and radial dust-gas coupling to study the anisotropy of the turbulence in the different rings. In order to be conservative, we compared the least constraining $\alpha_z/\St$ from \autoref{tab:verticalTrapping} to $\alpha_r/\St$ from \autoref{tab:radialTrapping}. Because most constraints are upper limits in the vertical direction, we also mostly get  upper limits on the ratio $\alpha_z/\alpha_r$. Moreover, we note that in some cases, the range of radii where the  radial constraints (centered on the rings) were obtained do not fully overlap with that of the vertical constraints (centered on gaps). To increase the number of rings with an indication for the anisotropy of turbulence, we decided to include these systems, however marking them by a $^+$  in \autoref{tab:radialTrapping} and with a lighter blue in Fig.~\ref{fig:radial_vertical_alpha}. The results are reported in the last column of \autoref{tab:radialTrapping} and shown in the third panel of Fig.~\ref{fig:radial_vertical_alpha}. 

We find that the upper limits for $\alpha_z/\alpha_r$ vary between 0.06 and 8.57, with a median value of about 1.60. Out of the 20 rings, 7 have $\alpha_z/\alpha_r \lesssim 1$, indicating that vertical turbulence is either weaker than radial turbulence within these rings~\citep[see also V4046 Sgr;][]{Weber_2022}. 
On the other hand, similarly to \cite{Doi_Kataoka_2023}, we find that vertical turbulence must be stronger than radial turbulence in the inner, vertically thick ring of HD163296 located around 67\,au. The other 12 rings do not allow to conclude on the anisotropy of turbulence and are consistent with  vertical turbulence being either comparable to or weaker than radial turbulence.

\subsection{Implications for the origin of turbulence}

Our analysis indicated that most of the radial features in our sample are compatible with either quasi-isotropic turbulence (i.e., $\alpha_z/\alpha_r\sim 1$) or with features being disproportionately wider in the radial direction compared to their vertical extent (i.e., $\alpha_z/\alpha_r<1$). Our estimates on the ratio $\alpha_z/\alpha_r$ can then be used to constrain the mechanism driving the measured turbulence levels.

Specifically, our results suggest that the vertical shear instability \citep[VSI,][]{Nelson_etal_2013} is incompatible with most features, as it has been shown to drive highly anisotropic turbulence that manifests as vertically elongated motions with a vertical extent significantly larger than its radial counterpart \citep{Nelson_etal_2013} and, consequently, $\alpha_z\gtrsim650\alpha_r\gg\alpha_r$ \citep{Stoll_2017}. It might still be possible for the VSI to be responsible for the vertical stirring of grains in some of the systems where our approach could not constrain $\alpha_r$, in which case a case-by-case analysis is necessary to examine whether the disk conditions can support the growth and saturation of the VSI at significant levels.

We note, however, that the VSI is commonly associated with an accretion stress or radial turbulence of $\alpha_r\sim10^{-5}$--$10^{-3}$ \citep[e.g.,][]{Stoll_etal_2014, Manger_etal_2020}, suggesting that a vertical turbulence
of $\alpha_z\gtrsim10^{-2}$--$10^{-1}$ is a reasonable expectation. Combined with typical Stokes numbers of $\St\sim10^{-3}$--$10^{-1}$ for millimeter-sized grains at $R\sim10 - 100$\,au scales (see also Sect.~\ref{sec:fragmentationlimit}, \autoref{tab:alpha}), we can narrow down the expected ratio $\alpha_z/\St$ to $\gtrsim0.1$ for typical disk parameters. This in turn can rule out the VSI as the driving mechanism for several features where our analysis showed that $\alpha_z/\St\lesssim0.1$ (highlighted in a light red region in the top panel of Fig.~\ref{fig:radial_vertical_alpha}), even if a direct measurement of $\alpha_z/\alpha_r$ was not possible. In addition, given that the strength of the VSI decays with increasing distance from the star as the dust--gas thermal coupling timescale renders cooling inefficient \citep[thereby quenching the VSI, see e.g.,][]{Pfeil_etal_2023}, we can argue that if a feature has been constrained to not be VSI-driven, any other features exterior to that in the same system are also unlikely to be VSI-related. 

Nevertheless, we can identify different turbulence-driving mechanisms that are compatible with our results. In particular, for $\St\lesssim10^{-1}$, the emergent turbulence due to the streaming instability \citep[SI,][]{Youdin_2005} tends to be isotropic between the radial and vertical directions even for an initial dust-to-gas ratio of order unity \citep[e.g.,][]{Johansen_Youdin_2007,Baronett_etal_2024}. Given that the majority of substructures analyzed here are compatible with radial dust trapping, which would naturally collect a high concentration of millimeter-sized grains with $\St\lesssim0.1$, we expect that the SI could be operating in several of these rings. This concept has indeed been explored by other studies that suggest using the observed azimuthal variation in intensity to further constrain the extent to which the SI is taking place \citep[e.g.,][]{Scardoni_2024}.

Another popular mechanism that is capable of both driving turbulence and creating the observed dust traps and substructure (rings, gaps) in the first place is, of course, embedded planets. The gravitational interaction between the planet and its surrounding disk gives rise to characteristic spiral waves \citep{Ogilvie_Lubow_2002}, which can deliver angular momentum into the disk \citep{Rafikov_2002}. Gap opening is then a natural byproduct of planet--disk interaction for even Saturn-mass planets in typical  ALMA targets \citep[more precisely, for a planet-to-star mass ratio of order $\sim (H_\mathrm{g}/r)^3$, see][] {Goodman_Rafikov_2001}, with dust being efficiently trapped at the resulting outer gap edge. Three-dimensional models have further shown that planets excite both radial and meridional motion at the gap edge and around the excited spiral waves, which can in turn both lift dust above the midplane \citep{Bi_etal_2021} and diffuse it radially about a dust trap \citep{Bi_etal_2023}. The resulting turbulent stresses $\alpha_r$ and $\alpha_z$ are of order $\sim10^{-3}$, suggesting that an assumption of $\alpha_z/\alpha_r\sim1$ is not an unreasonable expectation.

Finally, it is possible for the origin of turbulence to lie in magnetohydrodynamic effects such as the ambipolar diffusion-mediated magnetorotational instability \citep{Simon_etal_2013}, which manifests as isotropic turbulent motion, or to the gravitational instability \citep[GI,][]{Toomre_1964,Kratter_Lodato_2016}, which results in $\alpha_r>\alpha_z$ \citep{Bethune_etal_2021}. However, the former would require the presence of a non-negligible vertical magnetic field, the strength of which is difficult to constrain observationally \citep[e.g.,][]{Harrison_etal_2021}, while the high disk masses required and the lack of observed spiral structures make the GI highly incompatible with most systems in our sample.

Overall, we expect the VSI to be incompatible with most of the systems in our sample, as we typically find that $\alpha_z\lesssim\alpha_r$, while mechanisms such as the SI or planet--disk interaction are more favorable explanations for at least some of the observed substructures. Pinning down the dominant process that drives turbulence in each system would of course require a case-by-case analysis, which is beyond the scope of this study.

\section{Constraints on turbulence level and implications}
\label{sec:discussion}

\subsection{Estimating turbulence in the fragmentation limit}
\label{sec:fragmentationlimit}

In this section, we attempt to break the degeneracy between $\St$ and $\alpha$ by assuming that the maximum grain size is limited by fragmentation. This is motivated by the results from  laboratory experiments~\citep{Guttler_2010, Beitz_2011, Musiolik_2016}, which indicate that dust fragmentation can occur at low relative velocity, as low as tens of centimeter per second, depending on the dust composition. Observationally, multi-wavelengths observations have revealed relatively flat profiles for the maximum grain size as a function of radius in several disks~\citep[e.g.,][]{Sierra_2021, Guidi_2022}. Performing numerical simulations of dust evolution in protoplanetary disks, \citet{Jiang_2024} showed that this feature can be explained if dust growth is limited by fragmentation, with a fragmentation velocity of about 1\ ms$^{-1}$.

At the fragmentation barrier, the fragmentation threshold velocity $v_\text{frag}$ can be equated with an approximate turbulent collision speed, such that the turbulent velocity coefficient $\alpha_\text{frag}$ and the fragmentation Stokes number $\St_\text{frag}$ are related via the following~\citep[e.g.,][]{Birnstiel_2024}:
\begin{equation}
    \label{eq:eqfragmentation}
    \St_\text{frag} = \frac{1}{3\alpha_\text{frag}} \left(\frac{v_\text{frag}}{c_\text{s}} \right)^2.
\end{equation}
If the ratio $\zeta=\alpha/\St$ is known, \autoref{eq:eqfragmentation} can be rewritten as: 
\begin{equation}
    \label{eq:alphafrag}
    \alpha_\text{frag} = \frac{v_\text{frag}}{c_\text{s}}\sqrt{\frac{\zeta}{3}},\\
    \St_\text{frag} = \frac{v_\text{frag}}{c_\text{s}}\sqrt{\frac{1}{3\zeta}},
\end{equation} 
where $c_\text{s} = \sqrt{\frac{k_\text{B} T}{\mu m_\text{p}}}$ is the disk sound speed. 
Contrary to the classical estimate of the Stokes number (e.g., \autoref{eq:stokes}), the above formulations allow us to estimate the fragmentation turbulence level and Stokes number without any assumptions on the relevant dust size and gas density.

We estimated $\alpha_\text{frag}$ and $\St_\text{frag}$ using  \autoref{eq:alphafrag}. We assumed a fragmentation velocity of $v_\text{frag}=$ 1\ ms$^{-1}$, and used  $\zeta=\alpha_z/\St$; that is, the value for the vertical dust--gas coupling as constrained in Sect.~\ref{sec:vertical_turbulence} (\autoref{tab:verticalTrapping}). Effectively, this assumes that vertical turbulence is dominant over, or comparable to  radial turbulence. This is a reasonable assumption, as only a limited number of rings analyzed in Sect.~\ref{sec:anisotropyTurbulence} display a notably different relative turbulence strength (only 4/20 disks with $\alpha_z<0.2\alpha_r$, and 1 with $\alpha_z<0.1\alpha_r$). 
To estimate the disk sound speed, we considered the midplane temperature in the \texttt{mcfost} models, averaged over the radial domain of the constraint~(see Sect.~\ref{sec:vertical_turbulence}).   We report the results in \autoref{tab:alpha}, in addition to the assumed values in the \texttt{mcfost} models.

We find that the turbulence level obtained assuming the fragmentation limit  $\alpha_\text{frag}$ reaches upper limits between $1.7\times 10^{-4}$ and $5.8 \times10^{-3}$, with a median of $1.1\times 10^{-3}$ (\autoref{tab:alpha}). These estimates are almost systematically lower than their assumed values in the \texttt{mcfost} models, with an average difference of a factor about 2. Consequently, the Stokes numbers obtained assuming the fragmentation limit are also systematically lower than what is assumed in the \texttt{mcfost} models (see \autoref{tab:alpha}, Appendix~\ref{appdx:St_frag_vs_St_mcfost}), suggesting that the disk masses (or grain sizes) in \texttt{mcfost} may be too low (or too high) by a similar factor (see~\autoref{eq:stokes}). 

Five disks in our sample were previously included in the study of \cite{Jiang_2024}, who estimated $\alpha_\text{frag}$ for the same fragmentation velocity, using disk masses inferred from optically thin dust emission and a gas-to-dust ratio of 100, and considering the maximum dust size inferred from existing multi-wavelengths analysis. We find that our results are consistent with their findings as, in most cases, the values inferred by \cite{Jiang_2024} are lower than the upper limits obtained here. \cite{Zagaria_2023} used an alternative method to estimate the turbulence in HD163296 in the fragmentation limited regime. They used the observed dust sizes and midplane temperature obtained from multi-wavelengths observations combined with an estimate of $\alpha_r/\St$ to infer $\alpha_\text{frag}$. The turbulence value that they obtained for the outer ring is consistent with our estimate, while their lower limit for the turbulence in the inner ring is about 4 times larger than our estimate.

\subsection{Comparison with accretion turbulence }
\label{sec:accretionalpha}

Historically, turbulence was introduced as a way to explain the observational fact that disks accrete. Since the mechanism producing turbulence was not known at the time when disks were postulated to be turbulent, and its exact nature is still elusive many decades after, it has become common to adopt a single value for the turbulence strength $\alpha$, assuming this applies to the entire disk. While angular momentum transport and therefore accretion are influenced mostly by radial turbulence, it is also common to assume that turbulence is isotropic and therefore the radial and vertical values of $\alpha$ are the same.

In the previous sections however, we showed that the strength of turbulence can be different in the vertical and radial directions, with seven cases of  stronger radial turbulence than vertical turbulence, and 13 rings where the opposite cannot be excluded~(Fig.~\ref{fig:radial_vertical_alpha}). 
Now, we aim to compare the level of vertical turbulence $\alpha_\text{frag}$, which we estimated assuming that dust size is limited by fragmentation (Sect.~\ref{sec:fragmentationlimit}), to the level of turbulence which can be expected based on the accretion rate of the disks (hereafter accretion turbulence or $\alpha_\text{acc}$).  

Using disk global quantities, the accretion turbulence is related to the mass accretion rate, $\dot M$, and the total disk mass, $M_\text{disk}$, following~\citep[e.g.,][]{Rosotti_2023}:
\begin{equation}
    \alpha_\text{acc} = \frac{\dot M}{M_\text{disk}}\frac{\mu m_\text{p}}{k_\text{B} T} \Omega R_\text{c}^2,
    \label{eq:accretionAlpha}
\end{equation}
where $\mu$, $m_\text{p}$, $k_\text{B}$, $T$ have already been defined in Sect.~\ref{sec:vertical_turbulence}, $\Omega =\sqrt{GM_\star / R_\text{c}^3}$, and $R_\text{c}$ is a length-scale describing the disk size, that is, within which a significant fraction of its mass resides. This length-scale can be quantified using commonly used analytical functional forms for the disk surface density~\citep{LyndenBell_1974}. 

We remark that this formula is the most empirical estimate of accretion turbulence we can make; it does not prove in any way that disks are turbulent, but it quantifies how turbulent disks should be if accretion is indeed powered by turbulence. If this is not the case, the formula is still useful as it quantifies how effective angular momentum transport processes are. Finally, we note that we use global quantities here rather than local (e.g., disk mass rather than the disk surface density at the ring location) because those would require to make the further assumption that the disk is in steady state. Given the large radii of many of the rings we considered in this study, this is unlikely to be the case.

As with any estimate, the reliability of the values returned by \autoref{eq:accretionAlpha} depends on the reliability of the values used as input. Several attempts were made in the literature to estimate the accretion turbulence, starting from the seminal work of \citet{Hartmann1998}. One of the main variable in these studies is the assumed disk radius. For example \cite{Rafikov_2017} used the millimeter continuum dust sizes to estimate $\alpha_\text{acc}$, while \cite{Ansdell_2018} considered the observed $^{12}$CO size.  
As was stated above, the disk radius should instead be linked to where most of the disk mass is. Many works have shown that the dust radius fails to represent most of the gas mass due to radial drift;  
on the other hand, the $^{12}$CO size  is not a good indicator either due to the high CO optical depth, which leads to an overestimate of $R_\text{c}$. To ensure we have accurate estimates, we restrict our sample and consider here studies where the disk radius has been determined by two techniques. The first consists in fitting the disk rotation curve (\citealt{Martire_2024} for AS209, GM Aur, HD163296; and \citealt{Longarini_in_prep}, based on \citealt{Izquierdo_in_prep}, \citealt{Galloway_in_prep}, and \citealt{Stadler_in_prep} for LkCa 15),
which directly probes the disk radius through its mass. The second technique consists in using thermochemical models to derive the physical size from the observed $^{12}$CO size~\citep[][for the rest of the sample]{Trapman_2023}. 


The other variable to discuss is the disk mass. Dust based estimates are available for all our disks, but we opt not to use them as these are highly dependent on the assumed dust-to-gas ratio. Instead, we restrict the sample to disks where the gas mass has been estimated through the rotation curve, in the same way as the disk size  \citep[][for AS209, GM Aur, HD163296, LkCa 15]{Martire_2024, Longarini_in_prep}, or alternatively from CO gas line observations. In this latter case, since these estimates are dependent on the CO abundance, we only consider cases where the CO abundance could be inferred from combined N$_2$H$^+$ observations \citep[][for GW Lup and V1094Sco]{Trapman_inprep}, or safely assumed because the source is a Herbig star \citep[][for HD142666]{Stapper_2024}, for which the consensus is that they should have ISM-like levels of CO abundance. Finally, we note that we assume a disk temperature at $R_c$ based on our \texttt{mcfost} models.

\begin{figure}
    \centering
    \includegraphics[width = 0.47\textwidth]{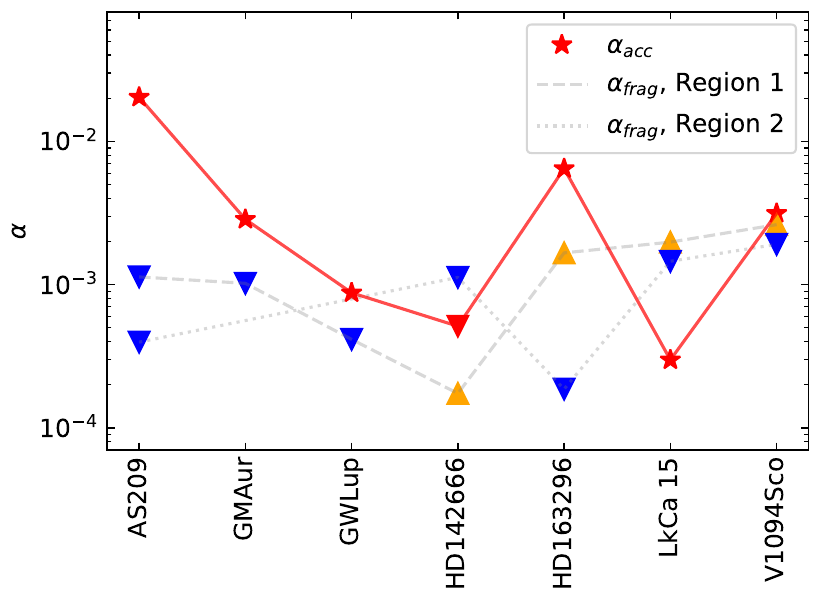}
    \caption{Comparison between accretion turbulence (in red) and the fragmentation turbulence $\alpha_\text{frag}$, derived from $\zeta=\alpha_z/\St$ (Sect.~\ref{sec:fragmentationlimit}).}
    \label{fig:alphafrag_alphaAcc}
\end{figure}

The results are shown in the top panel of Fig.~\ref{fig:alphafrag_alphaAcc}. We find accretion turbulence levels between $3.0\times 10^{-4}$ and $2.0\times 10^{-2}$, with a median of $3.0\times 10^{-3}$. 
Our values appear to be in line with the results of \citet{Ansdell_2018}. As was mentioned before, we note however that using the observed $^{12}$CO size the authors likely overestimated accretion turbulence. This could mean the sample we consider here is biased toward high accretors. Pending confirmation on a larger statistical sample where $\alpha_\text{acc}$ is estimated in a uniform and reliable way, if this is true this would make our sample a natural one where to search for turbulence and establish if it drives accretion.

Comparing $\alpha_\text{acc}$ with $\alpha_\text{frag}$ (Fig.~\ref{fig:alphafrag_alphaAcc}, \autoref{tab:alpha}), we find that, in 8 cases out of 14, the upper limits on turbulence we put in this paper are lower than what is required to drive accretion. At face value, this implies that in those disks turbulence is too small to explain the accretion rates we measure and therefore accretion is not driven by turbulence, or that the rings and gaps we studied in this work have a lower turbulence than the bulk of the disk. It should be remarked however that this is a strong statement only for AS~209, and region 2 of HD163296, where $\alpha_\text{frag}$ is significant lower than $\alpha_\text{acc}$ - the other cases are only a marginal mismatch. Conversely, for 6 cases out of 14, a mix of upper and lower limits, we find that our results are compatible with turbulence-driven accretion. In one case, region 1 of LkCa 15, the turbulence we detect is even too \textit{high} to explain accretion, implying that the local value we measure is probably significantly higher than in the rest of the disk. Overall, our results do not allow us to make strong conclusions regarding whether accretion is driven by turbulence at a population level, though it is significant that we find a few sources where turbulence seems unable to power accretion. 
We remark that at the population level this is a different outcome from a similar analysis in HL Tau \citep{Pinte_2016} and highlights the importance of conducting these studies for a sample.

\section{Implications for planet formation and limitations}
\label{sec:impplan}
\subsection{Implications for wide-orbit planet formation}

In the core accretion scenario, giant planets form by first accreting an approximately ten-Earth-mass solid core within the gas disk lifetime \citep{Mizuno_1980,Pollack_1996}. 
This is challenging in wide-orbits given the relatively short lifetimes of protoplanetary disks of a few million years, determined from the rapidly decreasing solid mass budget in nearby star-forming regions \citep{Ansdell_2017,Michel_2021,Manara_2023}. 
Despite this formation timescale challenge, wide-orbit giant planets are a relatively common type of planet with an approximate occurrence rate of between 5 and 10\% percent around solar-mass stars \citep{Nielsen_2019,Fulton_2021,Vigan_2021}. 
Moreover, the frequently observed dust rings in large protoplanetary disks, as explored in this work, could be consistent with a population of wide-orbit planets in the process of completing their growth \citep{Zhang_2018, Ruzza_2024}.

Core growth in wide orbits is maximally efficient when planetary embryos can accrete small pebbles \citep{Lambrechts_2012}, because small dust particles approximately millimeter in size experience gas drag which increases the accretion cross section \citep{Ormel_2010}. However, pebble accretion critically relies on pebbles settling toward the disk midplane. 
In order to assess which disk regions could be prone to drive core growth via pebble accretion, and which ones can crudely be ruled out, we can express a core formation time
\begin{align}
    t_{\rm acc,3D} = \frac{M_{\rm c}}{\dot M_{c,3D}},
\end{align}
where $\dot M_{c,3D}$ is the so-called 3D accretion rate onto an embryo of mass $M_{\rm c}$. This 3D accretion regime is valid when the accretion cross section is smaller than the pebble scale height $H_{\rm p}$ and is appropriate to describe the critical growth regime from planetesimal to larger embryos growing in the more efficient 2D pebble accretion regime. 

In this 3D growth phase, the accretion rate is proportional to the embryo mass, resulting in a mass-independent growth timescale that only depends on the particle Stokes number through the pebble disk aspect ratio ($H_{\rm p}/r$):
\begin{align}
    t_{\rm acc,3D} 
    &\approx  \eta \frac{H_{\rm p}}{r} F^{-1}  M_\star\,, 
\end{align}
for a given pebble flux, $F$ \citep{Lambrechts_2014}, and where $\eta$ here is a measure that expresses the pressure support in the disk \citep{Nakagawa_1986}. 

\begin{figure}
    \centering 
    \includegraphics[width = 0.47\textwidth]{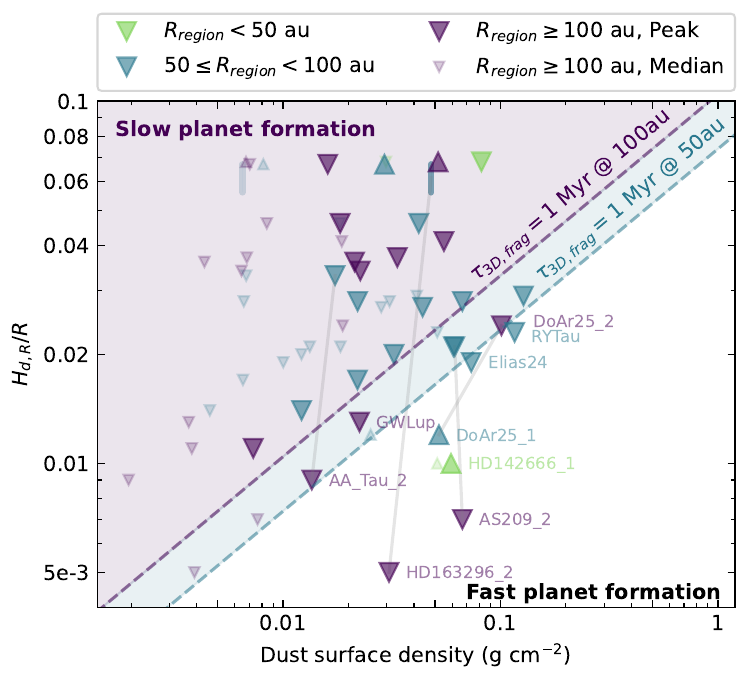}
    \caption{
    Critical dust surface density for core formation by pebble accretion as function of the solid surface density. 
    Large and small symbols indicate, respectively, the peak and median dust surface density for each of the modeled disk regions, while upper- and lower-limits on the dust disk aspect ratios are indicated by, respectively, downward and upward triangles. The ring location is color-coded according to orbital distance (see legend). The gray lines connecting triangles highlight two regions within the same disk. 
    The figure area below the dashed green  line marks rapid core growth timescales: below 1\,Myr at 50\,au in the 3D pebble accretion regime, assuming fragmentation-limited pebbles (see Sec.\,\ref{sec:impplan}). The dashed purple  line illustrates the 1-Myr-core-growth timescale at 100\,au. Disk regions with aspect ratios in the shaded purple  region are unlikely to form planetary cores within disk lifetimes.
    }
    \label{fig:growthTimescale}
\end{figure}

Making the same assumption as in Sec.\,\ref{sec:fragmentationlimit}, namely that pebbles are fragmentation limited, we can rewrite the pebble flux driven by radial drift in the Epstein regime, to obtain:
\begin{align}
    t_{\rm acc,3D}  
    &\approx
    \left( r \Sigma_{\rm p} \right)^{-1} 
    \left( \frac{H_p}{r} \right)^2
    v_{\rm frag}^{-1}
    M_\star,
     \label{eq:t3d}
\end{align}
where $\Sigma_{\rm p}$ is the pebble surface density and $v_{\rm frag}$ is the pebble fragmentation velocity. 
In this way, we obtain an expression where the pebble accretion timescale is fully determined by two observable quantities, the pebble aspect ratio and surface density. 
Finally, we can rewrite \autoref{eq:t3d} to express the critical pebble scale height needed for core growth timescales within disk lifetimes:
\begin{align}
  \left. \frac{H_p}{r} \right|_{\rm crit}  
  \approx & 
  \,0.033
  \left( \frac{r}{100\, \rm au} \right)^{1/2}
  \left( \frac{v_{\rm frag}}{1\,\rm ms^{-1}} \right)^{1/2}
  \left( \frac{t_{\rm acc, 3D}}{1\,\rm Myr} \right)^{1/2} \nonumber \\
  & \left( \frac{\Sigma_{\rm p}}{0.1\, \rm g/cm^2} \right)^{1/2}
  \left( \frac{M_\star}{M_\odot}\right)^{-1/2}\,.
   \label{eq:Hporcrit}
\end{align}
Here, we set the core growth timescale to $t_{\rm acc, 3D}$ to a characteristic disk age value of $t_{\rm acc, 3D}=1$\,Myr and assume solar-mass stars.  The dependency of the critical pebble disk aspect ratio on the local pebble surface density is illustrated in Fig.~\ref{fig:growthTimescale}.  

Several of the modeled disk regions have upper limits on the dust scale height that would allow core formation to proceed efficiently even at wide orbital distances outside of $50$\,au. 
Fig.~\ref{fig:growthTimescale} shows the observational constraints on the dust aspect ratio and dust surface density for all disks in this study (see Table~\ref{tab:results}) that can be compared with the critical pebble scale height of \autoref{eq:Hporcrit}. We show both the peak and median surface densities in each region, as obtained in Sect.~\ref {sec:fittingSurfaceDensity}, with respectively large and small triangle symbols.

In particular, the protoplanetary disk AS\,209 stands out: both ring at 74\,au and 120\,au (see Fig.~\ref{fig:finalCuts_AA_Tau_AS209_CI_Tau_DL_Tau_DS_Tau_Elias24}) have upper limits on the dust aspect ratio that place it in a region of parameter space where core growth by pebble accretion would be possible. 
Similarly, the disk DoAr\,25 (Fig.~\ref{fig:upperLower_DoAr25_HD142666}) contains two dust rings suitable for core growth by pebble accretion.
Finally, the low dust scale height in the outer ring of HD\,163296  (Fig.~\ref{fig:upperLower_LkCa15_V1094Sco}) makes it a region possibly prone to core growth even at $100$\,au distances.  
Interestingly, the above disks are also suspected to host wide-orbit planets, based on kinematic perturbations in CO emission: 
around 200\,au in  AS\,209 \citep{Bae_2022},
in the gap at $98$\,au around DoAr\,25 \citep{Pinte_2020}, and  
around 83 and 137\,au in HD\,163296 \citep{Teague_2018,Pinte_2018b, Izquierdo_2022}.
Therefore, it remains currently difficult to disentangle if these dust rings were the seeding grounds for core formation, or reflect current pile-up regions after giant planets already formed formed that are perhaps still prone to ongoing planet formation.

Some of the obtained lower-limits on dust scale heights already effectively rule out specific regions as sites for efficient core growth (upward triangles in the shaded purple area of Fig.~\ref{fig:growthTimescale}). 
Two exceptions to this are the lower-limits given for the inner rings in the disks of Doar\,25 and HD\,142666, the latter being the most close-in region probed in this study ($\lesssim$\,$50$\,au, Fig~\ref{fig:upperLower_DoAr25_HD142666}).  
In general, the current data is consistent with a possible trend of increased dust scale heights closer to the host star (bottom panel of Fig.~\ref{fig:alpha_constraints}, Sect.~\ref{sec:thickthin}). 
If this holds, the inner disk, crudely within half the dust radius, would be less conductive to core growth by pebble accretion. This is because the growth timescale expressed in \autoref{eq:t3d} mainly depends on the pebble scale height: the factor $r\Sigma_{\rm p}$ is nearly constant, if $\Sigma_{\rm p}$ approximately follows a nominal gas disk scaling $\Sigma_{\rm g} \propto r^{-1}$.  
In a broader perspective, this finding appears to be consistent with previous works proposing that giant planet formation is initiated in the outer disk, after which planets migrate inward, complete their growth and form the basis of the observed population of warm and cold Jupiter exoplanets. This has been argued on the basis of theoretical population synthesis models \citep{Lodato_2019,Ndugu_2019} and disk substructure/exoplanet statistical analysis \citep{vdmarel_2021}.

We have here presented an initial exploration to identify low pebble-scale-height disks with adequate pebble surface densities to be potential sites of pebble-accreting cores. Following works could target disks for more in-depth modeling and to verify that the fragmentation-limit assumed here holds by determining pebble sizes. This appears at least to be the case for AS\,209 and HD\,163296 \citep{Jiang_2024}.
A more detailed study could also explore the core formation process in more detail, fine-tuned per disk, by considering core growth from a pebble-dense planetesimal-forming region and subsequent inward migration. In that context, pebble accretion in pebble-rich rings can be highly efficient \citep{Jiang_2023}, unless a pressure-bump generated pebble ring contains only a limiting finite mass reservoir \citep{Morbidelli_2020}.

\subsection{Caveats and future prospects} 

This section discusses some caveats regarding the different aspects of this work, starting with observational considerations and moving toward more modeling related points. 
Some observational characteristics are not reproduced by our radiative transfer model. While these are located in regions where we do not put any constraints on the dust height, we point them out here because they could be interesting to study in future works. 
For example, several systems show deeper gaps along the minor axis than along the major axis direction (e.g., Elias27). In some cases (e.g., DL Tau, HL Tau, and V1094Sco) a ring which is well visible along the major axis direction completely disappears in the minor axis profile. In HL Tau (ring between 64-74au), this has previously been pointed out by \cite{Pinte_2016}. This effect is in appearance similar to the ring effect previously discussed in Sect.~\ref{sec:expectedEffects}. However, in all of these systems, even our thickest models are not able to reproduce the depth of the gap along the minor axis direction. For example, in the case of V1094Sco (Fig~\ref{fig:upperLower_LkCa15_V1094Sco}),  all models are similar between 164 and 201\,au and do not reproduce the gap in the data well. Further modeling would be required to fully understand the origins of this behavior. 

Moreover, a few systems (CI Tau,  ISO-Oph17, and J1610) show a surprising azimuthal variation in their brightness profiles, with the disk being brighter along the minor axis direction than along its major axis. This is not expected to be caused by projection effects and is not well reproduced by our radiative transfer models. We hypothesize that in some cases, these rings could be partially optically thick and that the effect detected here could be similar to that reported by~\cite{Scardoni_2024}.

We note here the degeneracy between the optical thickness and the true scale height. From the apparent vertical thickness, we assumed a reasonable optical thickness (surface density and dust properties) to estimate the dust scale height. However, we note that other combinations of optical thickness and dust scale height might be possible. 
Moreover, the dust heights reported in \autoref{tab:results} were obtained by fitting a Gaussian function to the vertical density profile of one dust size only, corresponding to dust with the maximum opacity at the wavelength of the observations (see Sect.~\ref{sec:relevantquantities}). We caveat however that the true vertical profile of the dust is not exactly a Gaussian (see \autoref{eq:fromang}), and that several dust sizes might be contributing to the final image resulting from radiative transfer modeling. 

For completeness, we also note that we assumed a constant dust-to-gas ratio radially throughout the disk. While we separate the inner and outer regions of the disk during our analysis, we note that the models do not include radially variable settling. We caveat that some effects of varying settling between the inner and outer regions, such as shadowing, cannot be captured under this assumptions. 
Furthermore, the settling prescription we assume supposes that the vertical  density profile of the gas follows a Gaussian, that diffusion is constant vertically, and that there is no feedback of dust to the gas. Regarding the latter point, \cite{Lim_2024} showed that the level of turbulence inferred by ignoring feedback (here, the model dependent $\alpha_{z, \text{MCFOST}}$) is potentially lower than that required in case where dust interacts with gas.

While we estimated the gas scale heights under the assumption of hydrostatic equilibrium (using \autoref{eq:Hg_temp}), direct observational constraints to the gas scale height would be valuable to better estimate the ratio $H_\text{d}/H_\text{g}$ and more precisely determine  the ratio $\alpha_z/\St$. This could for example be done by first determining the emitting surfaces of molecules~\citep[][]{Pinte_2018, Law_2021} and then retrieving the gas scale height using thermochemical modeling or analytical expressions~\citep[e.g.][]{Paneque-carreno_2023}. 

The estimate of $\alpha_r/\St$ is based on the assumption that the rings correspond to pressure maxima. However, only two systems in our sample have direct evidences that this is the case, with a direct measurement of the gas radial width ($w_\text{g}$) using molecular line emission~\citep{Rosotti_2020}. Expanding on the sample of rings with such direct estimates would also allow us to critically improve on our evaluation of $\alpha_r/\St$, and thus on the anisotropy of turbulence at various locations. This could be done with the observations of the exoALMA large program.  

Finally, the determination of $\alpha_\text{frag}$ depends on several assumptions which we state again here. First we assumed that the dust size is limited by fragmentation. This seems reasonable given the relatively low variation in the maximum grain size between the location of rings and gaps in different systems observed with multi-wavelengths observations~\citep{Sierra_2021, Jiang_2024} but might be incorrect in the outer disks where the grain size could be limited by radial drift instead. Further, we used our estimates of $\alpha_z/\St$ to obtain $\alpha_\text{frag}$, which effectively assumes that vertical turbulence is dominant, and we assumed a low fragmentation velocity of 1 ms$^{-1}$, as done in previous studies~\citep{Jiang_2024} and consistent with some laboratory experimental results~\citep{Guttler_2010}. In the case where the fragmentation velocity was larger or if vertical turbulence did not dominate we would expect the values of $\alpha_\text{frag}$ to become higher. This could affect the discussion of Sect.~\ref{sec:accretionalpha}, by changing the relative ratio between $\alpha_\text{frag}/\alpha_\text{acc}$ in some disks where $\alpha_\text{frag}$ is only marginally smaller than $\alpha_\text{acc}$.

\section{Conclusions}
\label{sec:conclusion}
 
We produced radiative transfer models of 33 inclined and ringed protoplanetary disks with high quality observations in the ALMA archive, aiming to constrain the vertical extent of the particles emitting in band 6 (1.3mm) or band 7 (0.9mm). Our analysis of the observations leads to the following results:
\begin{itemize}
    \item 10 disks do not allow us to provide significant constraints on the local dust scale height. 
    \item We provide upper limits on the dust scale height for the outer regions of the 23 remaining disks.  
    \item For 5 disks, we also obtain a lower limit on the dust scale height in the region within approximately half of the disk radius. Three of them (HD163296, LkCa 15, and V1094Sco), definitely exhibit a thicker aspect ratio in the inner ring  than in the outer disk.  
    \item By comparing the measured dust scale height with estimated gas scale heights, we evaluate the vertical ratio $\alpha_z/\St$. 
    We find that vertical settling is prominent in most systems, as is indicated by low values of $\alpha_z/\St<1$.
       \item    Three systems (HD163296, LkCa 15, and V1094Sco) showed evidence of enhanced mixing in their inner regions.
    \item We also estimate the radial ratio $\alpha_r/\St$ of several rings by comparing the locally estimated gas scale heights with the dust widths.  We find that $\alpha_r/\St<1$ for 11 rings, consistent with radial trapping, while the other 10 rings do not necessitate strong trapping. 
\end{itemize}

We explored the implications of our findings on dust settling and radial trapping for the anisotropy of turbulence, its origin, disk evolution, and planet formation. Our key results include:
\begin{itemize}
    \item The analysis of 20 rings reveals that vertical turbulence is typically comparable to or weaker than radial turbulence ($\alpha_z\lesssim\alpha_r$). The inner ring of HD163296 is a notable exception, displaying stronger radial than vertical turbulence, in agreement with previous studies. 
    \item The apparent turbulence anisotropy suggests that VSI is incompatible with most of the rings in our sample, as it produces strong vertical and weak radial turbulence. Mechanisms like the streaming instability or planet--disk interaction may better explain some of the observed substructures.   
    \item  Aiming to disentangle  $\alpha$ and $\St$, we assume that the dust size is limited by fragmentation, with a fragmentation velocity of 1 m\,s$^{-1}$. We use our estimate of $\alpha_z/\St$ to estimate $\alpha_\text{frag}$ and $\St_\text{frag}$ without assumption on the relevant dust size and gas density.  We find a median upper limit of $\alpha_\text{frag} \lesssim 1.1\times10^{-3}$.
    \item We compared $\alpha_\text{frag}$ to the turbulence level required for accretion in seven disks with well-constrained gas disk properties (gas size and gas mass). 
    Although we cannot make strong conclusions regarding the link between accretion and turbulence across all disks, we find few sources where the inferred turbulence cannot be the main source of accretion. 
    \item Finally, we analyzed our estimates for the dust scale height and dust surface densities in the context of pebble accretion. We find that several of the modeled disk regions have estimates that would allow core formation to proceed efficiently within about 1\,Myr, even at wide orbital distances outside of 50\,au.  In the case of the disks with vertically thicker inner than outer disk, we find that planet formation could be faster in the outer disk.
\end{itemize}

 This highlights the need for a better understanding of the physical structure of the disk and dust vertical settling across all radii to better comprehend the best locations for forming planets. 

\medskip
\small 
\noindent
\emph{Data availability:} 
The data can be found on zenodo (\href{https://doi.org/10.5281/zenodo.14054952}{zenodo.14054952}). The models are also available
(\href{https://doi.org/10.5281/zenodo.14056832}{zenodo.14056832}).

\medskip
\small 
\noindent
\emph{Acknowledgements:} 
We thank the referee for their review of the manuscript. 
The authors thank the DSHARP team and Jane Huang for making their calibrated visibilities and imaging scripts publicly available. We also thank Feng Long, Myriam Benisty, Jun Hashimoto, and \'Alvaro Ribas for sharing some of the calibrated visibilities used in this work, and the NRAO SRDP and ESO  ARI-L \citep{Massardi_2021} for providing calibrated visibility files. We thank Giuseppe Lodato for very useful discussions. 
MV and GR acknowledge support from the European Union (grant No. 101039651, project DiscEvol) and from Fondazione Cariplo (grant No. 2022-1217). FMe and GD acknowledge funding from the European Research Council (ERC) under the European Union's Horizon Europe research and innovation program (grant agreement No. 101053020, project Dust2Planets). AZ acknowledges funding by the STFC (grant ST/P000592/1), and from the European Union under the European Union's Horizon Europe Research and Innovation Programme 101124282 (EARLYBIRD). ML acknowledges ERC starting grant 101041466-EXODOSS. Views and opinions expressed are, however, those of the author(s) only and do not necessarily reflect those of the European Union or the European Research Council. Neither the European Union nor the granting authority can be held responsible for them.
This paper makes use of the following ALMA data: 2013.1.00498.S, 2015.1.00888.S, 2016.1.00484.L, 2016.1.00460.S, 2016.1.00545.S, 2016.1.00771.S, 2016.1.01164.S, 2016.1.01205.S, 2016.1.01239.S, 2016.1.01370.S,  2017.A.00006.S, 2017.1.01151.S,  2017.1.01167.S, 2017.1.01460.S, 2018.1.00028.S, 2018.A.00030.S, 2018.1.00689.S, 2018.1.00958.S, 2018.1.01255.S, 2018.1.01230.S, 2018.1.01755.S, 2019.1.01051.S. ALMA is a partnership of ESO (representing its member states), NSF (USA), and NINS (Japan), together with NRC (Canada), MOST and ASIAA (Taiwan), and KASI (Republic of Korea), in cooperation with the Republic of Chile. The Joint ALMA Observatory is operated by ESO, AUI/NRAO, and NAOJ. This paper used the following softwares \texttt{CASA}~\citep{CASA_2022}, \texttt{mcfost}~\citep{Pinte_2006, Pinte_2009}, \texttt{Matplotlib}~\citep{Hunter_2007}, \texttt{Numpy}~\citep{Harris_2020}, \texttt{scipy}~\citep{Scipy_2022}, \texttt{CARTA}~(\href{https://zenodo.org/doi/10.5281/zenodo.3377984}{10.5281/zenodo.3377984}). 

\normalsize

\bibliographystyle{aa.bst}
\bibliography{biblio}{}

\begin{appendix}
\FloatBarrier


\begin{table*}[h!]

\section{Tables}

 \FloatBarrier
 \begin{multicols}{2}
We present here the main tables including the results of this paper. Table~\ref{tab:disk_parameters} gathers the disk and stellar parameters considered for this study. Table~\ref{tab:results} presents the vertical constraints obtained in the different disk regions, the corresponding grain size for which they are valid, and the local and total dust mass in the models. Table~\ref{tab:verticalTrapping} shows the implications of these results for vertical trapping.  Table~\ref{tab:radialTrapping} indicates the level of radial trapping and turbulence anisotropy.  Finally, Table~\ref{tab:alpha} compares $\alpha_{\text{MCFOST}}$ assumed in the radiative transfer modeling with $\alpha_{frag}$ assumed based on the vertical constraints and assuming fragmentation limited dust (Sect~\ref{sec:fragmentationlimit}).  $\alpha_\text{acc}$ estimated to explain the level of accretion in a few systems (Sect.~\ref{sec:accretionalpha}) is also reported in Table~\ref{tab:alpha}.
\end{multicols}

 \FloatBarrier

\bigskip
    \centering
    \caption{Disk parameters assumed in our modeling.}
    \begin{tabular}{cccccccccccc}
    \hline\hline
    Name & Adopted & d  & i & PA & $R_{out}$ & $T_{eff}$ & $L_\star$& $R_\star$& $M_\star$ & $\dot M$ & References\\
    && (pc) & ($^\circ$) & ($^\circ$) & (au) & (K)  & ($L_\odot$) & ($R_\odot$) & ($M_\odot$) & ($M_\odot$/yr) & \\
    \hline 
    AA Tau & AA Tau & 137 & 59.0 & 93.0 & 160 & 4350 & 1.10 & 1.85 & 0.68 & $10^{-7.4}$& (20, 21, 23)\\
    AS209 & AS209 & 121 & 35.0 & 85.7 & 150 & 4265 & 1.40 &2.20 & 1.31& $10^{-7.3}$ & (1, 2, 34)\\
    CI Tau & CI Tau &160 & 49.2 & 11.3 & 300 & 4200 & 1.26 & 2.12 & 1.02 &  $10^{-8.2}$ &  (3, 4, 5, 6) \\
    DL Tau & DL Tau &159 & 45.0& 52.1 & 200 & 4277 & 0.64 & 1.47 & 0.98 & $10^{-7.2}$ & (7, 23)\\
    DoAr25 & DoAr25 & 138 & 67.0& 110.0 & 165 & 4266 & 0.95 &1.79 & 1.06 & $10^{-8.3}$ &(1, 2, 6)\\
    DS Tau & DS Tau &159 & 65.2 & 160.0 & 100 & 3792 & 0.25 & 1.15 & 0.58 & $10^{-7.3}$ & (7, 23)\\
    Elia 2-24 & Elias24 &136 & 29.0 & 45.7 & 160 & 4266 & 6.00 & 4.50 & 0.46 & $10^{-6.4}$ &(1, 2)\\
    GM Aur & GM Aur & 159 & 53.2 & 57.2 & 300 & 4350 &0.90 & 1.67& 1.32 & $10^{-8.0}$ & (8, 9, 10, 23)\\
    GO Tau & GO Tau &144 & 54.0& 21.0 & 150 & 3516 & 0.29 & 1.7* &0.36 & $-$& (7, 11)\\
    GW Lup & GWLup &155 & 38.7& 37.6 & 140 & 3630 & 0.33 & 4.0* & 0.46 & $10^{-9.0}$& (1, 2) \\
    FS Tau B & Haro 6-5B & 140 & 74.0 & 145.0 & 300 & 4265$^{g}$ & $-$ & 2.94$^{g}$ & $-$  & $-$ & (12)\\
    HD142666 & HD142666 &148 & 62.0 & 162.0 & 70 & 7586 & 9.12 & 1.74 & 1.73 & $<10^{-8.4}$ & (1, 2, 6)\\
    HD163296& HD163296& 101 & 46.7 & 133.0 & 169 & 9333 & 17.0 & 1.58 & 2.04 & $10^{-7.4}$ &(1, 2)\\
    HL Tau & HL Tau & 147 & 46.7 & 138.0 & 147 & 4000 & $-$& 40* & 1.70 &  $10^{-7.1}$ & (24, 25, 26)\\
    ISO-Oph 17 & ISO-Oph 17 & 140 & 42.4 & 131.0 & 84 & 3989 & 3.98 & 4.20 & 0.69 & $10^{-7.6}$ & (13, 14) \\
     J16083070-3828268& J1608 & 156 & 73.0 & 108.0 & 121 & 4800 & 3.02 & 2.00 & 1.40 & $10^{-9.2}$ & (15, 22, 23) \\
     J16100501-2132318& J1610 & 145 & 37.8 & 60.5 & 100 & 3950 & 0.46 &  1.45 & 0.67 & $-$& (16)\\
    LkCa 15 & LkCa 15 &159 & 50.2 & 61.9 & 250 & 4500 & 1.05 & 1.69 & 1.25  & $10^{-9.2}$ & (16, 17)\\
    MY Lup & MYLup & 156 & 73.0& 58.8 & 110 & 5129 & 0.87 & 1.18 &1.23 & $<10^{-9.6}$ &(1, 2) \\
     J16313124-2426281 & Oph163131 & 147 & 84.0 & 49.0 & 170 & 4500 & $-$ & 5.1* & 1.20 & $-$ &  (15, 18)\\
     PDS70 & PDS70 & 113 & 49.7 & 160.4 & 120 & 3972 & 0.35 & 1.26 & 0.76 &  ~$10^{-10.2}$ & (27, 28)\\
     RY Tau & RY Tau & 138 & 65.0 & 23.0 & 100 & 5945 & 11.8 & 3.25 & 1.95 &$10^{-7.7}$ & (29, 30)\\
    V1094Sco & V1094Sco &150 & 55.0 & 109.0 &300 & 4205 & 1.15 & 2.02 & 0.83 & $10^{-7.9}$ & (19, 14, 23)\\
    \hline
    \multicolumn{12}{c}{Disks without constraints on their vertical thickness}\\
    DoAr33 & DoAr33 & 139 & 41.8 & 81.1 & 27 & 4467 & 1.51 & 2.06 & 1.09 & $-$ & (1, 2)\\
    DM Tau & DM Tau & 145 & 35.2 & 157.8 & 120 & 3580 & 0.24 & 1.30 & 0.39 & $10^{-8.3}$ & (11, 32, 33)\\
    Elia 2-20 & Elias20 & 138 & 49.0 & 153.0 & 64 & 3890 & 2.23 &3.29 & 0.48 & $10^{-6.9}$ & (1, 2)\\
    Elia 2-27 & Elias27 & 116 & 56.2 & 119.0 & 254 & 3980 & 0.91 & 2.11 & 0.49 & $10^{-7.2}$ & (1, 2) \\
    HD143006 & HD143006 & 165 & 18.6 & 169.0 & 90 & 5623 & 3.80 & 2.10 & 1.78 & $10^{-8.1}$ & (1, 2)\\
    MWC 480 & MWC 480 & 147 & 36.5 & 147.5 & 135 & 8360 & 17.4 & 1.99 & 1.91 & $10^{-6.9}$ & (7, 31)\\
     J16255615–2420481  & SR4 & 134 & 22.0 & 18.0 & 35 & 4074 & 1.17 & 2.18 & 0.68 & $10^{-6.9}$ & (1, 2)\\
     Sz114 & Sz114 & 162 & 21.3 & 165.0 & 58 & 3162 & 0.20 & 1.51 & 0.17 & $10^{-9.1}$ & (1, 2)\\
    WaOph6 & WaOph6 & 123 & 47.3 & 174.2 & 103 & 4169 & 2.88 & 2.18 & 0.68 & $10^{-6.6}$ & (1, 2)\\
    WSB52 & WSB52 & 138 & 54.4 & 138.0 & 32 & 3715 & 0.70 & 2.03 & 0.48 & $10^{-7.6}$ & (1, 2)\\
    \hline
    \end{tabular}
    \\
    \tablefoot{ The bottom part of the table includes the sources for which the dust height could not be constrained.  In the first column ("Name"), we indicate a SIMBAD compatible name (2MASS to be added before the names starting with "J"), while the second column ("Adopted") shows the adopted name in the next tables. In the $R_\star$ column, the symbol $^{(*)}$ indicates that the reported $R_\star$ value is higher than the direct estimate from the literature. It must however not be considered as the true stellar radius as  a higher value was only introduced to ensure significantly optically thin/bright models (see Sect.~\ref{sec:fittingSurfaceDensity}). The symbol $^{(g)}$ in Haro6-5B implies that $T_{eff}$ and $R_\star$ were assumed because Haro6-5B does not have a direct measurement of these parameters and is a K5 star. }
    \tablebib{(1) \citet{Andrews_2018}; (2) \citet{Huang_2018}; (3) \cite{Clarke_2018}; (4) \cite{Gravity_collab_2023}; (5) \cite{Donati_2020}; (6) \cite{Law_2022}; (7) \cite{Long_2018}; (8) \cite{Huang_2020}; (9) \cite{Espaillat_2011}; (10) \cite{Macias_2018}; (11) \cite{Simon_2019}; (12) \cite{Villenave_2020}; (13) \cite{Cieza_2021}; (14) \cite{Testi_2022}; (15) \cite{Villenave_2019}; (16) \cite{Facchini_2020}; (17) \cite{Donati_2019}; (18) \cite{Villenave_2022}; (19) \cite{vanTerwisga_2018}; (20) \cite{Loomis_2017}; (21) \cite{Francis_2020}; (22) \cite{Alcala_2017}; (23) \cite{Manara_2023}; (24) \cite{Pinte_2016}; (25) \cite{Beck_2010}; (26) \cite{Stephens_2023}; (27) \cite{Keppler_2018}; (28) \cite{Haffert_2019}; (29) \cite{Ribas_2024}; (30) \cite{Mentiguia_2011}; (31) \cite{Mendigutia_2013}; (32) \cite{Kudo_2018}; (33) \cite{Hashimoto_2021}; (34) \cite{Martire_2024}} 
    \label{tab:disk_parameters}
    
 \FloatBarrier

\end{table*}

 \FloatBarrier

\begin{table*}
    \centering
     \caption{Constraints on the dust vertical concentration in different radial domains, local dust mass, and total dust mass. }
    \begin{tabular}{c|c|ccc|c|c|ccc}
    \hline\hline
    Name & $M_\text{dust}$ & Domain  & $H_{\text{d}, R}$& $H_{\text{d}, R}/R$  &Method &  $a_\text{MCFOST}$ & $\Sigma_\text{d, median}$ & $\Sigma_\text{d, max}$ \\
     & ($10^{-4}$ M$_\odot$) & (au)  & (au)  & ($-$) & ($-$)  & ($\mu$m) & ($10^{-3}$ g.cm$^{-2}$)  & ($10^{-3}$ g.cm$^{-2}$)\\
    \hline 
    AA Tau & $0.7 \pm 0.1$ &  43 $-$ 93& $\leq$ 2.3 & $\leq 0.033$& G &  88 & 6.8& 17.3 \\
        && ~~93  $-$ 160  & $\leq$ 1.2 & $\leq 0.009$ & O & 88  & 1.9 & 13.6\\
    AS209 & $2.6 \pm 0.3$& 39 $-$ 97  & $\leq$  1.4 & $\leq 0.021$ &R &  88 & 13.3 & 61.6\\
     &  & ~~97 $-$ 141   & $\leq$ 0.9 & $\leq 0.007$&  R & 88 & 7.6 & 66.8\\
    CI Tau & $4.2 \pm 1.0$ & ~~40 $-$ 130  &$\leq$ 3.9 & $\leq 0.046 $ & G/R & 88 & 18.3 & 42.1\\
    &&130 $-$ 234  &  ~~$\leq$  12.2 & $\leq 0.067$& R/O &88 & 7.0 & 16.1 \\
    DL Tau & $5.8 \pm 1.2$ & 46 $-$ 78  & $\leq$  1.7 & $\leq$ 0.028&G & 88 & 30.9 & 66.8\\
    DoAr25 & $6.4 \pm 1.7$ & 69 $-$ 86  & $\geq $ 0.9 & $\geq $ 0.012 &G & 88 & 25.3 & 52.1 \\
    && ~~86 $-$ 165 & $\leq $ 3.0 & $\leq 0.024 $& G/R/O  & 88 & 18.8 & 101.0\\
    DS Tau & $0.5 \pm 0.1$ & 55 $-$ 85  & $\leq $ 1.4 & $\leq 0.020$&O &88 & 12.1 & 32.4 \\
    Elias24 & $4.0 \pm 0.4$ & ~~37 $-$ 123 & $\leq 1.5$ & $\leq 0.019$&R & 88 & 10.0 & 73.4 \\ 
    GM Aur & $ 5.6 \pm 1.6 $& ~~40 $-$ 125  & $\leq$ 1.7 & $\leq 0.021$&G/R/O&  77 & 18.4 & 60.8 \\ 
    GO Tau & $1.7 \pm 0.5^*$ & ~~73 $-$ 150  & $\leq$ 4.1 &$\leq 0.037 $&G/R/O   & 88 & 6.8 & 33.5\\ 
    GW Lup & $1.2\pm 0.2^*$ & ~~62 $-$ 141  & $\leq$  1.3 &$\leq 0.013$&G/R/O & 88 & 3.7 & 22.5 \\ 
    Haro 6-5B & $3.5 \pm 0.9^g$&  110 $-$ 190 & $\leq$ 5.5 &$\leq 0.036$& O  & 59 & 4.3 & 21.3\\ 
    HD142666 & $0.9 \pm 0.1$& 11 $-$ 30  & $\geq$ 0.2 &$\geq 0.010$ & G  & 88 & 51.2 & 59.2\\ 
     &&40 $-$ 60  & $\leq$ 1.3 & $\leq 0.027$&O  & 88 & 28.3 & 43.8\\ 
    HD163296& $2.0 \pm 0.2$ & 49 $-$ 85 & 0.4 $-$ 4.5 & 0.006 $-$ 0.067 &R  & 88 & 6.5 & 48.0\\ 
      && ~~85 $-$ 155   & $\leq$ 0.6 & $\leq 0.005$&  R & 88 & 3.9 & 30.7\\ 
     HL Tau  & $3.2 \pm 0.5$ & ~~61 $-$ 150  & $\leq$ 5.5 & $\leq 0.053$& G/R/O &59 & 18.6 & 55.0\\
    ISO-Oph 17 & $1.5 \pm 0.2$ & 26 $-$ 46   & $\leq$ 2.4 & $\leq 0.068$& G  & 88 & 29.8 & 81.8\\ 
    J1608 & $0.2 \pm 0.0$& 60 $-$ 82 & $\leq$ 1.0 & $\leq 0.014$&R &88&  4.6 & 12.2\\ 
    J1610 & $0.2 \pm 0.0$ & 29 $-$ 78 & $\leq$ 1.5 & $\leq 0.028$&G/O& 88 & 6.6 & 22.0\\ 
    LkCa 15 & $2.1\pm 0.3$& 47 $-$ 89  & $\geq$ 4.5 & $\geq 0.067$& R& 88 & 8,1 & 29.3 \\ 
    && ~~89 $-$ 200 & $\leq$ 4.9 & $\leq 0.034$& R &88 &6.4 & 22.7\\ 
    MYLup & $2.8 \pm 0.8$& ~~40 $-$ 100 & $\leq$ 2.0 & $\leq 0.029$&O  & 88 & 41.3 & 127.7 \\
    Oph163131 & $0.6 \pm 0.1^*$& 107 $-$ 150 & $\leq $ 1.4 & $\leq 0.011$&O  & 88 & 3.8 & 7.3 \\
    PDS70 & $0.5 \pm 0.0$ & ~~50 $-$ 105  & $\leq$ 1.3 &  $\leq$ 0.017& R/O & 59 & 6.5 & 22.0\\
    RY Tau & $2.1 \pm 0.4$ & $28 - 90$ & $\leq$ 1.4 & $\leq$ 0.023 & O & 88 & 51.2 & 116.3 \\
    V1094Sco & $6.2 \pm 1.3$ & 107 $-$ 164  & $\geq$ 8.9 & $\geq 0.068$ & R& 88 & 6.7 & 51.7\\
    && 201 $-$ 280  & ~~$\leq$ 10.7 & $\leq 0.046$& R/O& 88 & 8.4 & 18.3\\
    \hline
    HD163296 & &67.9& $\geq3.8$ & $\geq0.056$& (1)&&\\
    \hline
    \end{tabular}
    \tablefoot{ In the "Method" column, G indicates that the gap contrast was determinant to put the constraint, O marks that the outer edge effect was mostly used, and R indicates that it is the azimuthal variation in the ring that was critical.  The seventh column $a_\text{MCFOST}$ indicates the grain size which has the maximum opacity at the respective wavelength of the observations and for which the constraints on $H_{d,R}$ and $H_{d, R}/R$ were obtained. The last two columns indicate the median and maximum surface density of each region. In the M$_{dust}$ column, indicating the total dust mass, $^{(*)}$ and  $^{(g)}$ highlight sources where the stellar radius was modified to lead to optically thinner dust emission (see Sect.~\ref{sec:fittingSurfaceDensity}, \autoref{tab:disk_parameters}), potentially affecting the dust mass estimate. }
    \tablebib{(1) \cite{Doi_Kataoka_2021}. } 
    \label{tab:results}
\end{table*}

\begin{table*}
    \centering
    \caption{Vertical trapping. }
    
    \begin{small}
    \begin{tabular}{c|ccc|cc}
    \hline\hline
    Name & Domain  &  $H_\text{g, model}$ & $\alpha_z/\St|_\text{model}$& $H_\text{g, temp}$ & $\alpha_z/\St|_\text{temp}$\\ 
     &(au) & (au) &  ($-$) &(au)& ($-$) \\ 
    \hline 
    AA Tau & 43 $-$ 93 & 6.4 & $\leq$ 1.5e-1 & 5.6 & $\leq$ 2.0e-1 \\
    & ~~93  $-$ 160 &  12.6~~ & $\leq$ 9.1e-3 &  10.8~~ & $\leq $ 1.3e-2\\ 
    AS209 &  39 $-$ 97 & 6.1 & $\leq$  5.5e-2 & 3.3 & $\leq$ 2.2e-1 \\
    & ~~97 $-$ 141 & 12.0~~ & $\leq$ 5.7e-3 & 6.4 & $\leq $ 2.0e-2\\
    CI Tau & ~~40 $-$ 130 & 7.5  &$\leq$ 3.7e-1 &  5.0 & $\leq$ 1.6~~~~~  \\
    & 130 $-$ 234 &18.8~~ &  $\leq$  7.3e-1 & 14.6~~ & $\leq$ 2.3~~~~~\\ 
    DL Tau & 46 $-$ 78 & 5.8 & $\leq$  9.5e-2  & 3.2 & $\leq$ 3.9e-1\\
    DoAr25 & 69 $-$ 86 & 7.5 & $\geq $ 1.5e-2  & 4.2 & $\geq $ 4.7e-2 \\
    & ~~86 $-$ 165 & 12.4~~ & $\leq $ 6.2e-2 & 7.6 & $\leq  $ 1.8e-1 \\
    DS Tau & 55 $-$ 85 & 6.6 & $\leq $ 4.6e-2 & 4.8 & $\leq$ 9.2e-2\\
    Elias24 & ~~37 $-$ 123 & 5.2 & $\leq $ 9.3e-2  & 5.4 & $\leq $ 8.3e-2\\
    GM Aur & ~~40 $-$ 125 & 7.3 & $\leq$ 5.7e-2& 4.3 & $\leq $ 1.9e-1\\
    GO Tau & ~~73 $-$ 150 & 10.8~~  & $\leq$ 1.7e-1 & 10.3~~ &$\leq$ 1.9e-1 \\
    GW Lup & ~~62 $-$ 141 & 9.6 & $\leq$  1.8e-2 & 9.2~~ &$\leq $ 2.0e-2\\
    Haro 6-5B & 110 $-$ 190 & 15.2~~  & $\leq$ 1.5e-1  &$-$&$-$\\
    HD142666 & 11 $-$ 30 & 1.6  &  $\geq$ 1.6e-2 & 0.9 &$\geq $ 5.6e-2 \\
    & 40 $-$ 60 & 4.5 & $\leq$ 8.7e-2 & 2.6 & $\leq $ 3.4e-1\\
    HD163296& 49 $-$ 85 &  6.3 & 5.8e-1 $-$ 1.06 & 3.9 & $-$\\
    & ~~85 $-$ 155 & 11.9 & $\leq$ 2.6e-3 & 7.2 & $\leq $ 7.0e-3\\
    HL Tau & ~~61 $-$ 150 & 10.1~~ & $\leq$ 2.2e-1 & 9.3 & $\leq$ 2.7e-1\\
    ISO-Oph 17 & 26 $-$ 46 & 3.2  & $\leq$ 1.4~~~~~  & 2.5 & $\leq$ 8.5~~~~~ \\ 
    J1608 & 60 $-$ 82 & 6.8  & $\leq$ 2.2e-2 & 6.4 & $\leq $ 2.5e-2 \\
    J1610 & 29 $-$ 78 & 4.6 & $\leq$ 1.2e-1 & 3.7 & $\leq $ 1.9e-1 \\
    LkCa 15 & 47 $-$ 89 & 6.3 & $\geq$ 1.0~~~~~ & 4.7 & $\geq $ 9.1~~~~~ \\
    & ~~89 $-$ 200 & 14.2~~ & $\leq$ 1.4e-1 & 10.5~~ & $\leq$ 2.8e-1\\
    MYLup & ~~40 $-$ 100 & 6.3 & $\leq$ 1.1e-1  & 3.2 & $\leq$ 6.3e-1\\
    Oph163131 & 107 $-$ 150 & 13.0~~ & $\leq $ 1.2e-2 & 10.3~~ & $\leq $ 1.8e-2\\
    PDS70 & ~~50 $-$ 105 & 7.2 & $\leq$ 3.3e-2 & 6.2 & $\leq$ 4.6e-2 \\
    RY Tau & 28 $-$ 90 & 5.0 & $\leq$ 8.4e-2 & 2.8 & $\leq$ 3.5e-1\\
    V1094Sco & 107 $-$ 164 & 13.8~~ & $\geq$ 7.2e-1 & 11.5~~ & $\geq $ 1.5~~~~~ \\
    & 201 $-$ 280 & 26.0~~ & $\leq$ 2.0e-1 & 22.5~~ & $\leq $ 2.9e-1\\
    \hline
    \end{tabular}
    \end{small}

\tablefoot{  The gas scale height  corresponds either to that assumed in the parametrization of the model ($H_\text{g, model}$, from \autoref{eq:Hg}) or to the scale height obtained based on the hydrostatic equilibrium ($H_\text{g, temp}$, from \autoref{eq:Hg_temp}), using the midplane temperature of the model. A $\alpha_z/\St$ ratio is estimated for both assumptions.}
    \label{tab:verticalTrapping}
    
 \medskip

    \centering
    \caption{Radial trapping and anisotropy of turbulence. }
   \begin{small}
    \begin{tabular}{c|ccccc|l}
    \hline
    Name &  Domain & $w_\text{d}$ & $w_\text{g, min}$ & $w_\text{g, max}$ & $\alpha_r/\St$&$\alpha_z / \alpha_r$ \\
     &(au) & (au)  &(au) & (au) &  \\
    \hline \hline
    AA Tau &~~73 $-$ 113 & $4.2\pm0.1$ & 7.8 & 11.4 & (0.14, 0.40) & $\leq 1.47^+$ \\ 
    AS209 & 64 $-$ 84 & $2.5\pm 0.1$ & ~~~7.4$^{(3)}$ & ~8.6$^{(3)}$   & (0.09, 0.13) &$\leq 2.46$\\ 
         & 110 $-$ 130 & $3.4 \pm 0.1$ & ~10.5$^{(3)}$ & 11.9$^{(3)}$  & (0.08, 0.12) & $\leq 0.25$\\
    CI Tau & 132 $-$ 192 & $22.6\pm 0.4$~~ & 13.1* & 28.1 & (1.90, ...) & $\leq 1.22$\\
    DL Tau & 70 $-$ 86 & $4.0\pm 0.1$ & 4.4 & 9.4 & (0.23, 5.31) & $\leq 1.74^+$\\
        & 107 $-$ 125 & $5.7\pm 0.1$ & 7.0 & 22.9 & (0.07, 1.97) & ~~~~$-$\\
    DoAr25 & 107 $-$ 115 & $2.9\pm 0.1$ & 6.8 & 10.7 & (0.07, 0.23) & $\leq 2.60$\\
        & 128 $-$ 148 & $3.5 \pm 0.8$ & 8.9 & 10.1 & (0.08, 0.18) & $\leq 2.40$\\
    DS Tau & 40 $-$ 70 & $6.8\pm 0.2$ & ~~3.6* & 19.1 & (0.13, ...) & $\leq 0.68^+$\\
    Elias24 & 67 $-$ 87 & $3.5\pm 0.1$ & 7.5 & 15.4 & (0.05, 0.28) & $\leq 1.73^+$\\
    GM Aur & 72 $-$ 96 & $5.2 \pm 0.1$ & ~~5.2* & 13.2 & (0.18, ...) & $\leq 1.04$\\
    GO Tau & 61 $-$ 85 & $4.4\pm 0.1$ & 6.4 & 14.8 & (0.10, 0.95) & $\leq1.90^+$ \\
        & 106 $-$ 122 & $5.8\pm 0.1$ & 11.5~~ & 18.7 & (0.10, 0.34) & $\leq 1.80$\\
    GW Lup & $79-89$ & 3.1$^{(1)}$ & 8.0 & 9.3 & (0.13, 0.18) & $\leq 0.16$\\
    Haro6-5B & 105 $-$ 115 & $5.6 \pm 0.2$ & $-$ & 6.9 & (1.56, ...) & $\leq0.10^+$\\ 
    HD163296 & 58 $-$ 76 & $4.3 \pm 0.1$ & ~13.4$^{(3)}$ & 15.4$^{(3)}$  & (0.08, 0.12) & $\geq 7.20$\\
        & ~~95 $-$ 105 & $3.0 \pm 0.1$  &~21.9$^{(3)}$ & 24.5$^{(3)}$    & (0.01, 0.02) & $\leq 0.54$\\
    J1610 & 24 $-$ 34 & $2.7 \pm 0.1$& ~~2.0* & 18.0 & (0.02, ...) & $\leq 8.57^+$\\
        &35 $-$ 47 & $3.9 \pm 0.1$ & ~~3.0* & 4.8 & (1.47, ...) & $\leq 0.13$\\
    LkCa 15 & 59 $-$ 79 & 6.3$^{(2)}$ & ~~5.1* & 18.4 & (0.13, ...) & ~~~$ 7.63^\star$\\
    Oph163131 & ~~94 $-$ 120 & $9.8 \pm 2.6$ & 8.9* & 14.6 & (0.32, ...) & $\leq 0.06^+$\\
    \hline
    \end{tabular}
    \end{small}
    \tablefoot{$w_\text{g, min}$ was obtained from the hydrostatic equilibrium using the midplane temperature of the model. The upper limit on the pressure bump width, $w_\text{g, max}$, was estimated from the separation of the peak of the ring to the nearest minimum. We mark rings where  $w_\text{d} > w_\text{g, min}$ with the symbol $^{(*)}$ in the $w_\text{g, min}$ column. The last columns shows the anisotropy of turbulence (see Sect.~\ref{sec:anisotropyTurbulence}). Some rings are not fully included in the domain of the vertical constraint (centered on gap, \autoref{tab:verticalTrapping}) and are marked with $^+$. The ratio between a lower limit in $\alpha_z/\St$ and in $\alpha_r/\St$ is marked with $^\star$ (LkCa 15), and is only informative. }
   \tablebib{ (1) \cite{Jennings_2022}, (2) \cite{Facchini_2020}, (3) \cite{Rosotti_2020}
    }
    \label{tab:radialTrapping}
\end{table*}

\begin{table*}
    \centering
    \caption{$\alpha$ and $\St$ assumed in the models, derived based on fragmentation limited dust sizes, or to explain accretion. }
    \begin{tabular}{c|ccc|cc|c}
    \hline\hline
    Name &  Domain  & $\alpha_{z, \text{MCFOST}}$ & $\St_\text{MCFOST}$ & $\alpha_\text{frag}$ & $\St_\text{frag}$ & $\alpha_\text{acc}$ \\
     & (au) & ($-$)  & ($\times10^{-3}$)  & ($\times10^{-4}$)  & ($\times10^{-3}$)  & ($\times10^{-4}$) \\
    \hline 
    AA Tau &  43 $-$ 93&  $\leq$ 2e-2 & 35.0 &$\leq$ 10.3 &$\geq $ 5.1 & $-$\\
        & ~~93  $-$ 160  &$\leq$ 2e-3 &  85.0 &$\leq$ 3.2 &$\geq $ 25.6 &$-$ \\ 
    AS209 & 39 $-$ 97 &  $\leq$ 2e-3  &  10.6 &$\leq$ 11.4 &$\geq $ 5.1 & 203.3\\
     & ~~97 $-$ 141   & $\leq$ 2e-4 & 19.9 &$\leq$ 4.0 &$\geq $ 19.8 & 203.3\\
    CI Tau &  ~~40 $-$ 130  & $\leq$ 2e-2  & 14.9 & $\leq$ 34.7 &$\geq $ 2.2  & $-$\\
    &130 $-$ 234  & $\leq $ 2e-1  & 61.4 &$\leq$ 51.7 &$\geq $ 2.2  & $-$\\ 
    DL Tau & 46 $-$ 78  & $\leq$ 2e-3 & 8.3&$\leq$ 17.4 &$\geq $ 4.4 & $-$\\
    DoAr25 &  69 $-$ 86  & $\geq $ 2e-4 &  7.2 &$\geq $ 3.3 & $\leq$22.9 & $-$\\
    & ~~86 $-$ 165  & $\leq$ 2e-3 & 11.4 &$\leq$ 13.6 &$\geq $ 7.4& $-$\\
    DS Tau  & 55 $-$ 85  & $\leq $ 2e-3  & 18.5 &$\leq$ 8.9 &$\geq $ 9.7 & $-$\\
    Elias24 & ~~37 $-$ 123  & $\leq$ 2e-3 & 10.4 &$\leq$ 6.5 &$\geq $ 7.0 & $-$\\
    GM Aur & ~~40 $-$ 125 & $\leq$ 2e-3& 9.9 &$\leq$ 10.2 &$\geq $ 5.4 & 28.5\\
    GO Tau &  ~~73 $-$ 150& $\leq$ 2e-2  & 36.0 &$\leq$ 14.9 &$\geq $ 8.0 & $-$\\
    GW Lup & ~~62 $-$ 141 & $\leq$ 2e-3 & 39.2 &$\leq$ 4.1 &$\geq $ 20.4 & 8.8\\
    Haro 6-5B&  110 $-$ 190& $\leq$ 2e-2 & 37.2 &$\leq$ 10.5 &$\geq $ 7.0 & $-$\\
    HD142666 & 11 $-$ 30  & $\geq$ 2e-4  & 6.1 &$\geq $ 1.7 &$\leq$ 11.1 & $\leq$ 5.1\\
     &40 $-$ 60 & $\leq$ 2e-3 & 11.2 &$\leq$ 11.2 &$\geq $ 3.3 & $\leq$ 5.1\\
    HD163296&  49 $-$ 85& 2e-4 $-$ 2e-1 & 18.8 & 16.4 & 2.0 & 64.4\\
      & ~~85 $-$ 155  & $\leq$ 2e-4 & 31.7 &$\leq$ 1.9 & 26.6& 64.4\\
     HL Tau& ~~61 $-$ 150 & $\leq$ 2e-2 & 10.5 &$\leq$ 8.3 &$\geq $ 3.1 & $-$\\
    ISO-Oph 17 & 26 $-$ 46   & $\leq$ 2e-1 & 10.3 &$\leq$ 58.4 &$\geq $ 0.7 & $-$\\
    J1608 & 60 $-$ 82 & $\leq$ 2e-3  & 55.7 &$\leq$ 2.2 &$\geq $ 9.0 & $-$\\
    J1610 & 29 $-$ 78 & $\leq$ 2e-2 & 31.0 &$\leq$ 10.3 &$\geq $ 5.4 & $-$\\ 
    LkCa 15 & 47 $-$ 89 & $\geq$ 2e-1 & 35.3&$\geq $ 19.8 &$\leq$ 1.9 & 3.0\\
    & ~~89 $-$ 200  & $\leq$ 2e-2  & 35.7 &$\leq$ 14.5 &$\geq $ 5.3 & 3.0\\
    MYLup & ~~40 $-$ 100  & $\leq$ 2e-3  & 4.2 &$\leq$ 22.1 & $\geq $3.5 & $-$\\
    Oph163131 & 107 $-$ 150 & $\leq $ 2e-3 & 87.2 &$\leq$ 3.0 &$\geq $ 17.0 & $-$\\
    PDS70 &  ~~50 $-$ 105  & $\leq$ 2e-3 & 24.6 &$\leq$ 4.8 &$\geq $ 10.5 & $-$\\
    RY Tau &$28 - 90$ &  $\leq$ 2e-3 & 3.7 &$\leq$ 11.3 &$\geq $ 3.2 & $-$\\
    V1094Sco  & 107 $-$ 164 & $\geq$ 2e-1 & 39.2 &$\geq $ 26.2 &$\leq$ 3.6 & 31.4\\
    & 201 $-$ 280 & $\leq$ 2e-2 & 42.8 &$\leq$ 19.1 &$\geq $ 6.5 & 31.4\\
    \hline
    \end{tabular}
    \tablefoot{Leftmost columns, $\alpha_z$ and Stokes numbers assumed in our \texttt{mcfost} modeling for a grain size $a_\text{MCFOST}$ indicated in \autoref{tab:results}. The middle columns assume dust size is limited by fragmentation, with a fragmentation velocity of $v_\text{frag}=1$ m\,s$^{-1}$, using the most conservative values of $\zeta=\alpha_z/\St$ from \autoref{tab:verticalTrapping} (see Sect.~\ref{sec:fragmentationlimit}). The rightmost column show the estimated global accretion turbulence for disks discussed in Sect.~\ref{sec:accretionalpha}. We note that the ratio $\alpha_{z, \text{MCFOST}}/\St_\text{MCFOST}$ are slightly different from the values reported in \autoref{tab:verticalTrapping}, possibly because the constraints on $H_{\text{d}, R}$ actually originate from the contribution of several grain sizes instead of just that with the maximum opacity.} 
    \label{tab:alpha}
\end{table*}

\clearpage

 \FloatBarrier

\begin{table*}

\section{Observational parameters}
\label{app:observations}

 \FloatBarrier

 \begin{multicols}{2}

We present the characteristics of the archival observations that we have used for this work. We have modeled a total of 33 protoplanetary disks. In \autoref{tab:observations}, we report the project IDs, beam size and wavelengths of the observations of the 23 sources where some constraints on their level of vertical turbulence could be obtained. On the other hand, \autoref{tab:observations_no_constraints} shows the same characteristics for the other disks modeled during this project but where no constraints have been obtained. When no constraints could be obtained, the disks were either not very inclined, possessed only shallow substructures, or the presence of spiral interior to the gap of interest made the interpretation of the model incorrect.

\end{multicols}
 \FloatBarrier
 
 \bigskip

    \centering
    \caption{Sample and image properties of the disks with constraints on their vertical thickness. }
    \begin{tabular}{cccccccccc}
    \hline
    Name & ALMA Project IDs & {Robust}&Beam & {rms}& References& $\lambda$ \\
    & &  & (\arcsec) & (Jy/beam) & &(mm) \\
    \hline \hline
    AA Tau & 2016.1.01205.S  & 0.5 & 0.095 $\times$ 0.061 & $2.5\times 10^{-5}$ & (1, 2, 3)& 1.25\\
    AS209 & 2016.1.00484.L  & -0.5~ & 0.038 $\times$ 0.036 & $1.7\times 10^{-5}$ &(4) & 1.25 \\
    CI Tau & 2016.1.01370.S, 2016.1.01164.S  & 0.5 & 0.046 $\times$ 0.032 & $1.4\times 10^{-5}$ &(5, 6)& 1.25\\ 
     DL Tau & 2016.1.01164.S & 0.5 & 0.134 $\times$ 0.109 & $4.8\times 10^{-5}$ &(6) & 1.25\\
     DoAr25 & 2016.1.00484.L  & 0&0.041 $\times$ 0.021 & $1.3\times 10^{-5}$ & (4) & 1.25\\
     DS Tau & 2016.1.01164.S & 0.5 & 0.144 $\times$ 0.099 & $4.9\times 10^{-5}$ & (6) & 1.25\\
     Elias24 &  2016.1.00484.L& 0 & 0.038 $\times$ 0.034 & $1.9\times 10^{-5}$ &(4)  & 1.25  \\
     GM Aur & 2017.1.01151.S, 2018.1.01230.S & 0.5 & 0.045 $\times $ 0.025 & $1.0\times 10^{-5}$ & (7)& 1.10\\
     GO Tau & 2016.1.01164.S & 0.5 & 0.140 $\times$ 0.107 & $4.7\times 10^{-5}$ & (6) & 1.33\\
    GW Lup &  2016.1.00484.L  & 0.5 & 0.045 $\times$ 0.043 & $1.5 \times 10^{-5}$ & (4)  & 1.25\\
    Haro 6-5B &  2016.1.00460.S  & 0 & 0.097 $\times$ 0.061 & $10.2\times 10^{-5}$~~ & (8)& 0.89\\
    HD142666& 2016.1.00484.L & 0.5 & 0.031 $\times$ 0.022 & $1.3\times 10^{-5}$ &  (4)  & 1.25\\ 
    HD163296& 2016.1.00484.L & -0.5~~ & 0.048 $\times$ 0.038 & $2.3\times 10^{-5}$ &  (4)  & 1.25\\ 
     HL Tau & 2019.1.01051.S & 0.5 & 0.035 $\times $ 0.030 & $2.4\times 10^{-5}$ & (15) & 0.89\\
     ISO-Oph 17 & 2016.1.00545.S, 2018.1.00028.S & 0.5 & 0.026 $\times$ 0.023 & $2.7\times 10^{-5}$ & (9, 10) & 1.25\\ 
    J1608 & 2018.1.00689.S & 0.5 & 0.020 $\times$ 0.019 & $1.5\times 10^{-5}$ &  $-$ & 1.25\\
    J1610 & 2018.1.01255.S & 0.3 & ~~~0.048 $\times$ 0.043$^{(a)}$ & $1.5\times 10^{-5}$ &  (11) & 1.25 \\
    LkCa 15 & 2018.1.01255.S & 0 & 0.051 $\times$ 0.034 &$1.2\times 10^{-5}$ &  (11)& 1.25\\
   MY Lup &  2016.1.00484.L  & 0 & 0.044 $\times$ 0.043 & $1.6\times 10^{-5}$ & (4)& 1.25 \\
     Oph163131 & 2016.1.00771.S, 2018.1.00958.S & 0.5 & 0.025 $\times$ 0.020 & $0.9\times 10^{-5}$ & (12, 13) & 1.33\\
    PDS70 & 2015.1.00888.S, 2017.A.00006.S,  & 0.5 & 0.038 $\times$ 0.035 & $1.3\times 10^{-5}$ & (16)& 0.89\\
    & 2018.A.00030.S\\
    RY Tau & 2016.1.01164.S, 2017.1.01460.S  &  0.5 & 0.047 $\times$ 0.030 & $3.6\times 10^{-5}$ & (17)& 1.33\\
    V1094Sco & 2016.1.01239.S, 2017.1.01167.S & 0.5 & 0.142 $\times$ 0.131 & $3.6\times 10^{-5}$ & (14) & 1.25\\
    \hline
    \end{tabular}
    \\
    \tablefoot{$^{(a)}$ For J1610, we used only the data observed in 2019, previously published by \cite{Facchini_2020}, and not the data observed in 2021. }
    \tablebib{(1) \cite{Loomis_2017}, (2) Loomis et al. in preparation, (3) \cite{Francis_2020}, (4) \cite{Andrews_2018}, (5) \cite{Clarke_2018}, (6) \cite{Long_2018}, (7) \cite{Huang_2020}, (8) \cite{Villenave_2020}, (9) \cite{Cieza_2019}, (10) \cite{Cieza_2021}, (11) \cite{Facchini_2020}, (12) \cite{Flores_2021}, (13) \cite{Villenave_2022}, (14) \cite{vanTerwisga_2018}, (15) \cite{Stephens_2023}, (16) \cite{Benisty_2021}, (17) \cite{Ribas_2024}}
    \label{tab:observations}

\bigskip
    \centering
    \caption{Sample and image properties of the disks without constraints on their vertical thickness. }
    \begin{tabular}{cccccccccc}
    \hline\hline
    Name  & ALMA Project IDs & Robust &Beam &rms &References &$\lambda$\\
    & && (GHz) &  (Jy/beam) && (\arcsec)  \\
    \hline 
    Elias20 & 2016.1.00484.L & 0& 0.032 $\times$ 0.023 & $1.5\times 10^{-5}$ &  (1) & 1.25\\ 
    Elias27 & 2016.1.00484.L  & 0.5 & 0.049 $\times$ 0.047 & $1.5\times 10^{-5}$ & (1)& 1.25\\
    DoAr33 & 2016.1.00484.L  & 0 & 0.037 $\times$ 0.024 & $1.7\times 10^{-5}$ & (1)& 1.25\\ 
    DMTau & 2013.1.00498.S, 2018.1.01755.S & 0.5 & 0.032 $\times$ 0.021 & $0.8\times 10^{-5}$ & (2)& 1.25\\ 
    HD143006 & 2016.1.00484.L & 0 & 0.053 $\times$ 0.036 &$1.1\times 10^{-5}$ &  (1)& 1.25\\  
    MWC 480 & 2016.1.01164.S  & 0.5& 0.166 $\times$ 0.110 & $6.2\times 10^{-5}$ & (3)& 1.25\\ 
    SR4 & 2016.1.00484.L  & -0.5~ & 0.053 $\times$ 0.036 & $2.5\times 10^{-5}$ & (1)& 1.25\\ 
    Sz114 & 2016.1.00484.L  & 0.5 & 0.066 $\times$ 0.028 & $1.9\times 10^{-5}$ & (1)& 1.25\\ 
    WaOph6 & 2016.1.00484.L  & 0 & 0.058 $\times$ 0.054 & $1.8\times 10^{-5}$ & (1)& 1.25\\ 
    WSB52 & 2016.1.00484.L & 0 & 0.033 $\times$ 0.027 & $1.6\times 10^{-5}$ & (1)& 1.25\\ 
        \hline
    \end{tabular}
    \\
	\tablebib{(1) \cite{Andrews_2018}, (2) \cite{Hashimoto_2021}, (3) \cite{Long_2018}}
    \label{tab:observations_no_constraints}
\end{table*}

 \FloatBarrier

\clearpage

\section{Comparison of $\St_\text{frag}$ with $\St_\text{MCFOST}$}
\label{appdx:St_frag_vs_St_mcfost}

\begin{figure}[h]
    \centering
    \includegraphics[width = 0.45\textwidth]{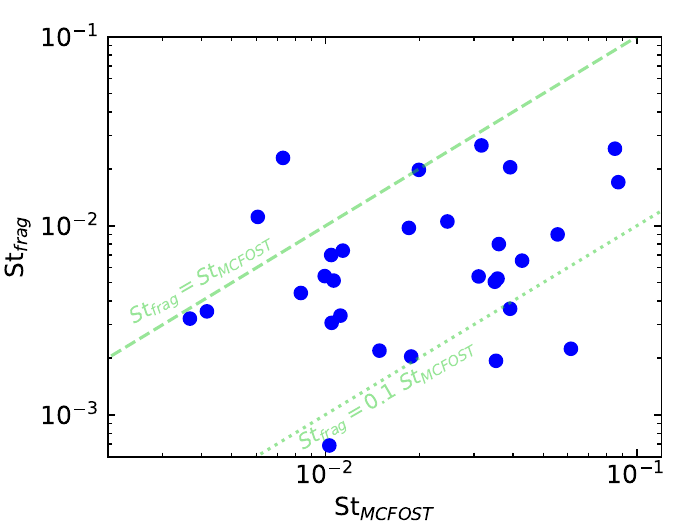}
    \caption{Comparison of $\St_\text{frag}$ (Sect.~\ref{sec:fragmentationlimit}) with $\St_\text{MCFOST}$ (Sect.~\ref{sec:relevantquantities}).}
    \label{fig:St_frag_vs_St_mcfost}
\end{figure}

In Fig.~\ref{fig:St_frag_vs_St_mcfost}, we show how the assumed Stokes number in our \texttt{mcfost} models compares with the estimated Stokes number assuming grain growth is limited by fragmentation. We assumed a fragmentation velocity of 1 m s$^{-1}$ and used the observational constraints on $\zeta=\alpha_z/\St$ to obtain $\St_\text{frag}$ (see Sect.~\ref{sec:fragmentationlimit}). $\St_\text{frag}$ is nearly systematically lower than $\St_\text{MCFOST}$ suggesting that the disk masses obtained with our iterative fitting of the surface density in \texttt{mcfost} may be too low.

\section{Cuts and metric estimates}
\label{app:cuts}

\begin{figure}
    \centering
    \includegraphics[width = 0.5\textwidth, trim={1cm 0.5cm 13cm 0cm},clip]{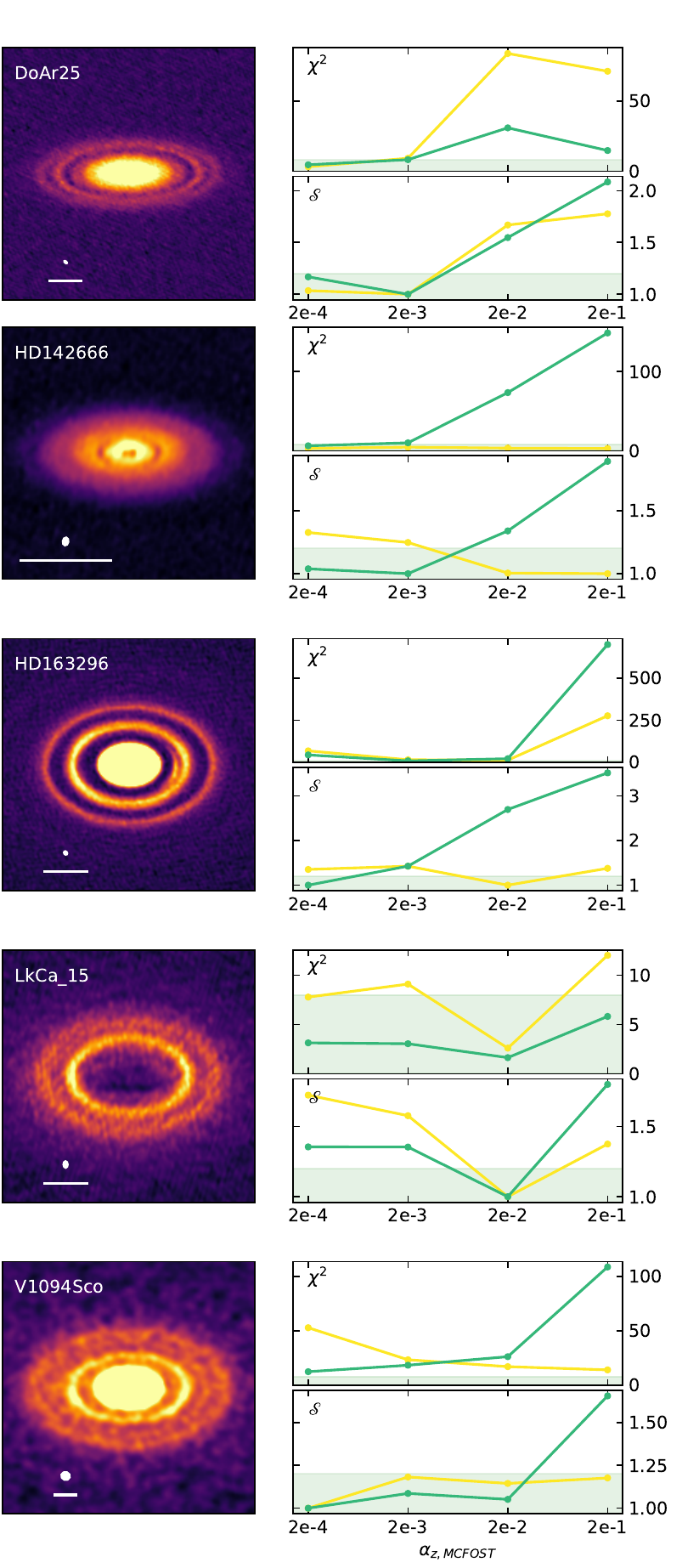}
    \caption{\emph{Left:} Data image, \emph{Right:} Values for the $\chi^2$ and shape parameter $\mathscr{S}$ (Sect.~\ref{sec:bestmodel_def}), quantifying the quality of the models. The green regions indicate that the models are generally good models. Yellow curves indicate results for the inner ring/region while green curves show that for the second ring/region. Major axis profiles highlighting the radial location of the different regions are shown in Fig.~\ref{fig:upperLower_LkCa15_V1094Sco} and Fig.~\ref{fig:upperLower_DoAr25_HD142666}.  The beam size and a 25\,au scale are indicated in the bottom left corner of the first panels.}
    \label{fig:finalCuts_DoAr25_HD142666_HD163296_LkCa_15_V1094Sco}
\end{figure}

\begin{figure*}
    \centering
    \includegraphics[width = 0.965\textwidth, trim={0cm 1.cm 0cm 0.6cm},clip]{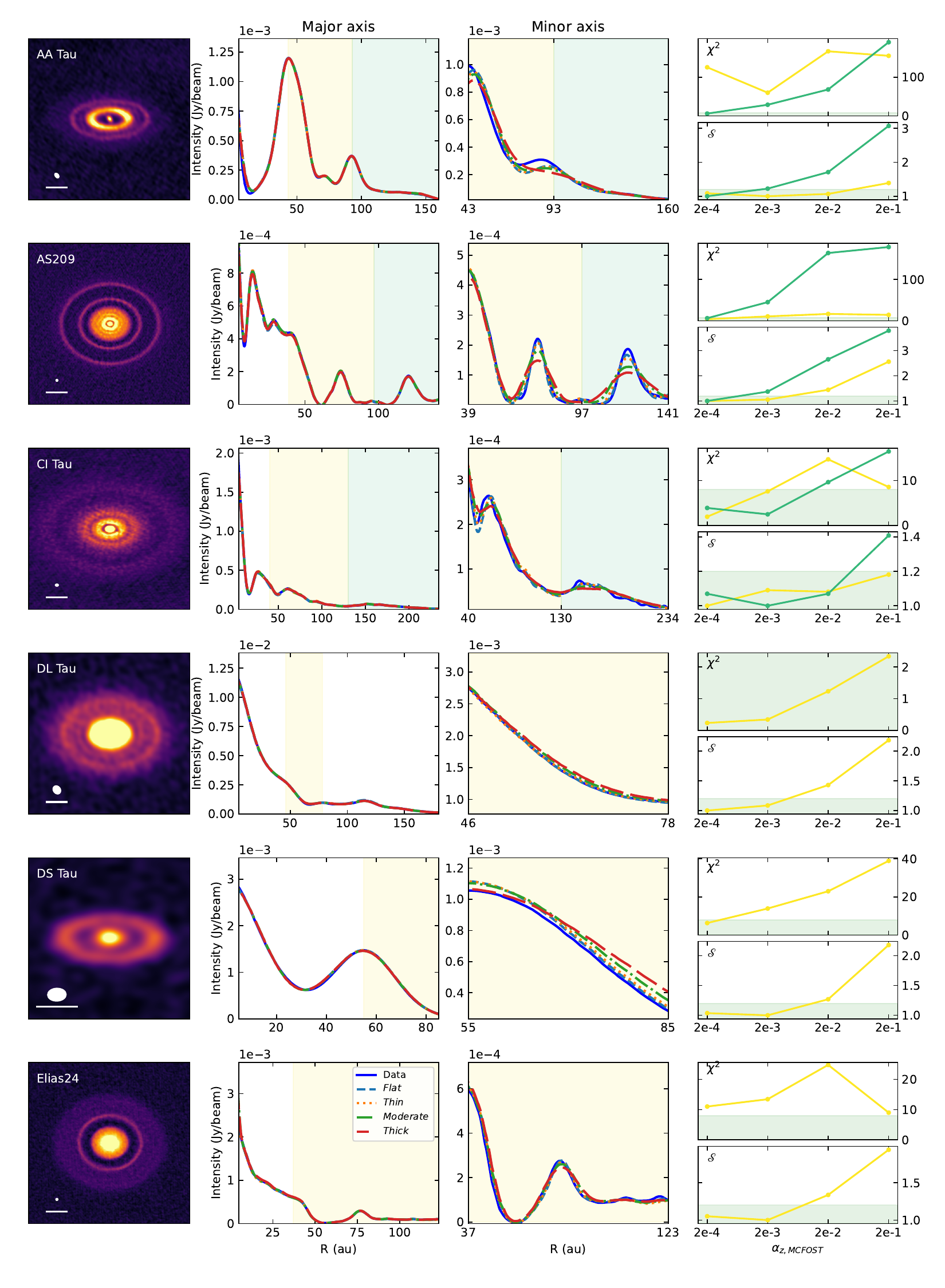}
    \caption{\emph{Left:} Data image, \emph{Middle left:} Major axis profiles of the data and models outside of 5au. \emph{Middle right:} Profiles along an azimuth of the data and model, zoomed in the constrained region. \emph{Right:} Values for the normalized $\chi^2$ (top) and shape parameter $\mathscr{S}$ (bottom), introduced in Sect.~\ref{sec:bestmodel_def}, quantifying the quality of the models. The green regions indicate that the models are generally good models. The colors of the curves correspond to the  similarly colored range of radius in the major axis profile displayed in the second left panel.  The beam size and a 25\,au scale are indicated in the bottom left corner of the first panels.}
    \label{fig:finalCuts_AA_Tau_AS209_CI_Tau_DL_Tau_DS_Tau_Elias24}
\end{figure*}

\begin{figure*}
    \centering
    \includegraphics[width = \textwidth, trim={0cm 1cm 0cm 0.6cm},clip]{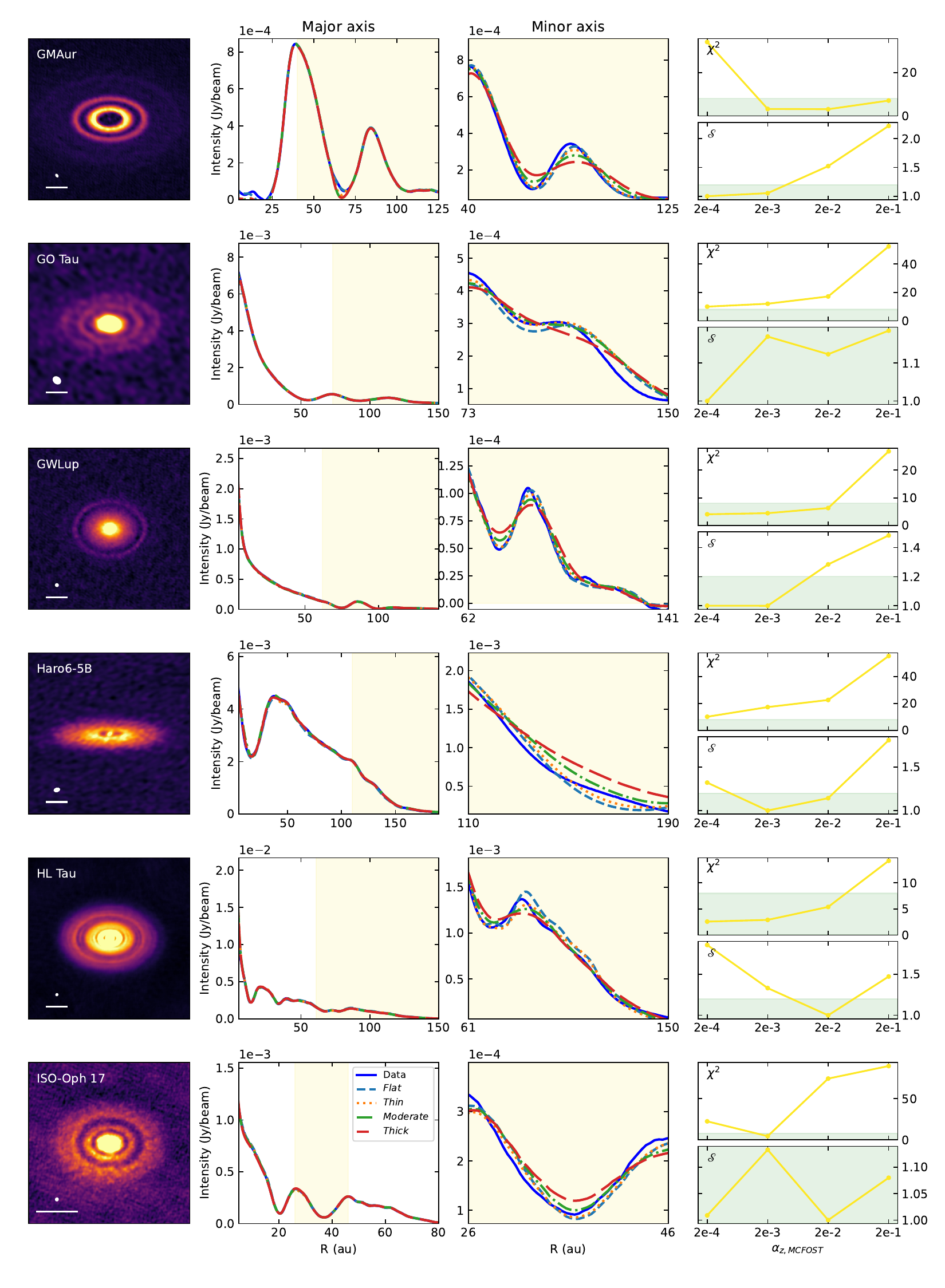}
    \caption{Same as Fig.~\ref{fig:finalCuts_AA_Tau_AS209_CI_Tau_DL_Tau_DS_Tau_Elias24}. For GM Aur, $\chi^2$  is artificially higher for model 2e-4 because $\chi^2_{majorAxis}$ is significantly better than for the other models.}
    \label{fig:finalCuts_GMAur_GO_Tau_GWLup_Haro6}
\end{figure*}

\begin{figure*}
    \centering
    \includegraphics[width = \textwidth, trim={0cm 0.5cm 0cm 0.6cm},clip]{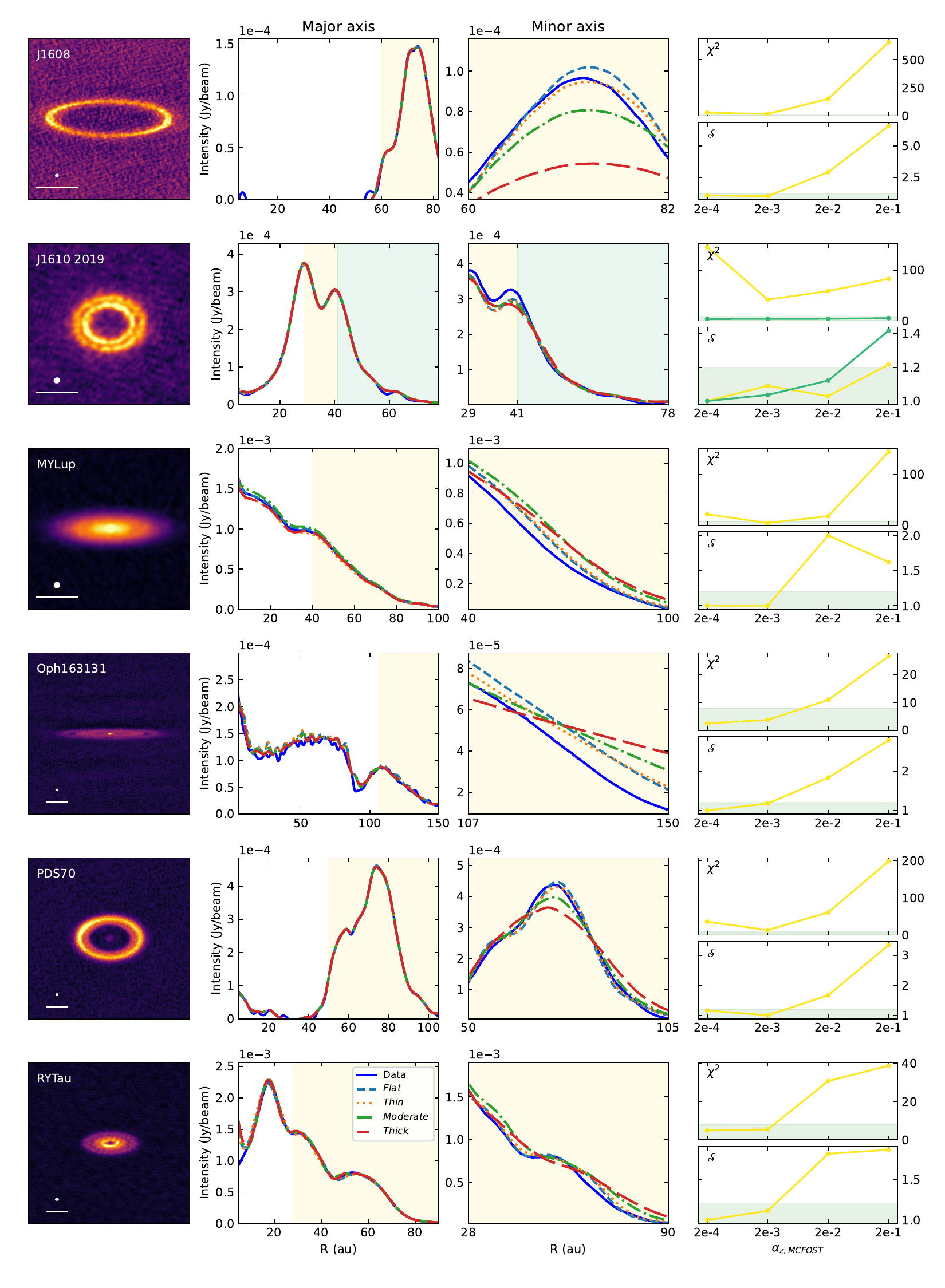}
    \vspace{-1cm}
    \caption{Same as Fig.~\ref{fig:finalCuts_AA_Tau_AS209_CI_Tau_DL_Tau_DS_Tau_Elias24}.}
    \label{fig:finalCuts_J1608_J1610_2019_MYLup_Oph163131_PDS70_RYTau}
\end{figure*}

In Fig.~\ref{fig:finalCuts_DoAr25_HD142666_HD163296_LkCa_15_V1094Sco}, we show images and metrics for the five disks for which we find that the inner disk half shows a finite dust thickness while the outer disk is settled (Sect.~\ref{sec:thickthin}). 
In Fig.~\ref{fig:finalCuts_AA_Tau_AS209_CI_Tau_DL_Tau_DS_Tau_Elias24}, Fig.~\ref{fig:finalCuts_GMAur_GO_Tau_GWLup_Haro6}, and Fig.~\ref{fig:finalCuts_J1608_J1610_2019_MYLup_Oph163131_PDS70_RYTau}, we present an overview of the 18 disks for which only upper limits on their vertical thickness could be obtained (Sect.~\ref{sec:upperlimits}). 
For all models, we also show the values of $\chi^2$ and $\mathscr{S}$ (Sect.~\ref{sec:bestmodel_def}).

We remind the reader that the normalized $\chi^2$ corresponds to the average of  $\chi^2_{azimuth} / \chi^2_{majorAxis}$ over each azimuthal direction. On the other hand $\mathscr{S}$, estimates the Hausdorff distance, which takes into account the likelihood in shape between the model and the data.  By visual inspection, we identify that models with $\chi^2 \leq 8$ and $\mathscr{S} \leq 1.2$ are generally good models. However, we always perform a visual check before reporting the value in \autoref{tab:results} and in some rare cases, the reported values might be slightly different from those highlighted by the metrics.

 \end{appendix}

\end{document}